\documentclass[12pt,a4paper]{article}

\usepackage{amsmath}
\usepackage{amssymb}
\usepackage{graphicx}
\usepackage{dsfont}
\usepackage{overpic}

\let\oldappendix=\appendix
\let\oldsection=\section
\renewcommand{\appendix}{\oldappendix%
\def\theequation{\Alph{section}.\arabic{equation}}%
\renewcommand{\section}{\setcounter{equation}{0}\oldsection}}

\newcommand{\beq}{\begin{equation}}
\newcommand{\eeq}{\end{equation}}
\newcommand{\beqa}{\begin{eqnarray}}
\newcommand{\eeqa}{\end{eqnarray}}
\newcommand{\no}{\nonumber}
\newcommand{\q}{\quad}
\newcommand{\qq}{\qquad}

\newcommand{\sfrac}[2]{{\textstyle\frac{#1}{#2}}}
\newcommand{\pn}{\tilde{\pi}^0}
\newcommand{\deltaph}{\lambda}
\newcommand{\newop}[2]{\def#1{\mathop{\mathrm{#2}}\nolimits}}
\newop{\artanh}{artanh}

\newcommand{\mcirc}[2][2]{{\overset{\circ}{m}\vphantom{m}_{#2}^{#1}}}

\newcommand{\mtop}[1][2]{\mcirc[#1]{0}}
\newcommand{\tad}[1]{\Delta_{#1}}
\newcommand{\tsum}{\mathop{\textstyle\sum}}
\newcommand{\cder}{D}
\newcommand{\decay}{f}
\newcommand{\coeffv}[2]{v_{#1}^{(#2)}}
\newcommand{\cbeta}[2]{\beta_{#1}^{(#2)}}
\newcommand{\cvtwid}[2]{\tilde{v}_{#1}^{(#2)}}

\newcommand{\GeV}{\,\mathrm{GeV}}
\newcommand{\Lagr}{\mathcal{L}}

\begin{document}

\hfill 

\hfill 

\bigskip\bigskip

\begin{center}

{{\Large\bf  Two-photon decays of $\mbox{\boldmath$\pi^0$}$, $\mbox{\boldmath$\eta$}$
  and $\mbox{\boldmath$\eta'$}$}}

\end{center}

\vspace{.4in}

\begin{center}
{\large B. Borasoy\footnote{email: borasoy@ph.tum.de},
 R. Ni{\ss}ler\footnote{email: rnissler@ph.tum.de}}

\bigskip

\bigskip

Physik Department \\
Technische Universit{\"a}t M{\"u}nchen \\
D-85747 Garching, Germany \\

\vspace{.2in}

\end{center}

\vspace{.7in}

\thispagestyle{empty} 

\begin{abstract}
We investigate the decays of $\pi^0, \eta$ and $\eta'$ into two photons
in an effective $U(3)$ chiral 
Lagrangian approach without employing large $N_c$ arguments.
Tree level and one-loop contributions from the anomalous Wess-Zumino-Witten
Lagrangian are calculated and the importance of $\eta$-$\eta'$ mixing
is discussed.
Unitarity corrections beyond one-loop  play an important role for the decays
with off-shell photons and are included by employing a coupled channel
Bethe-Salpeter equation which satisfies unitarity constraints
and generates vector-mesons from composed states of two pseudoscalar mesons.

\end{abstract}\bigskip

\begin{center}
\begin{tabular}{ll}
\textbf{PACS:}&12.39.Fe, 13.40.Hq\\[6pt]
\textbf{Keywords:}& Electromagnetic decays, chiral Lagrangians, unitarity, \\
 & resonances.
\end{tabular}
\end{center}


\vfill

\section{Introduction}\label{sec:intro}
The two-photon decays $\pi^0, \eta, \eta' \rightarrow \gamma \gamma$
are phenomenological manifestations of the chiral anomaly of QCD and
can provide important information on the chiral symmetry of the strong 
interactions. In order to study the phenomenological
implications of the chiral anomaly, it is convenient
to employ a chiral effective Lagrangian, 
since at low energies the effective degrees of freedom
are colorless hadrons rather than quarks and gluons. The effective Lagrangian contains a piece 
which reproduces the anomalous behavior of the effective action under chiral
transformations. Such a Lagrangian was systematically constructed by Wess
and Zumino \cite{WZ} by directly integrating the anomalous Ward identities
and later Witten provided a representation of the anomaly which illustrated
the topological content of the theory \cite{W}.

There are many anomalous processes which can be calculated from the Wess-Zumino-Witten (WZW)
Lagrangian, such as $\pi^0 \rightarrow \gamma \gamma, \eta \rightarrow \gamma \gamma,
\gamma \rightarrow 3 \pi, \eta \rightarrow \pi^+ \pi^- \gamma$ etc.
Originally, the WZW Lagrangian was formulated for the eight Goldstone bosons
$(\pi, K, \eta)$ which form an octet under chiral $SU(3)$ transformations,
but it can be easily extended to include the $\eta'$, the singlet counterpart of the
Goldstone boson octet \cite{DHL, BBC1, KL1, H-S2}.
Although the $\eta'$ is not a Goldstone boson due to the axial $U(1)$ anomaly
of the strong interactions, it combines with the Goldstone bosons to a nonet
at the level of the effective theory. This allows the phenomenological
investigation of $\eta'$ decays, such as $\eta' \rightarrow \gamma \gamma$
and $\eta' \rightarrow \pi^+ \pi^- \gamma$.

The present work deals with the two-photon decays of $\pi^0, \eta$ and $\eta'$
with one, both or none of the photons being off-shell, $\pi^0, \eta, \eta' 
\rightarrow \gamma^{(*)} \gamma^{(*)}$.
The decays $P \rightarrow \gamma \gamma \; (P=\pi^0, \eta, \eta') $
with both photons on-shell
are dominated by the WZW Lagrangian. They have been calculated up to one-loop order
within chiral perturbation theory (ChPT) and were shown not to receive non-analytic
corrections from the one-loop diagrams which were compensated by the chiral
corrections of the pseudoscalar decay constants \cite{DHL, BBC1}. This is also in agreement with
the complete one-loop renormalization of the anomalous Lagrangian \cite{DW, BBC2, Bij},
where it was shown by using heat kernel techniques that in dimensional regularization
one-loop diagrams for $P \rightarrow \gamma \gamma$ do not lead to divergences
that must be renormalized by appropriate counterterms of higher chiral order.
It was furthermore argued \cite{DHL} that a consistent picture of $\eta$-$\eta'$
mixing emerged from the two-photon decays with one mixing angle of approximately
$-20^\circ$.

More recently, a two-mixing angle scheme has been proposed by Kaiser
and Leutwyler 
\cite{KL1,L,KL2} for the calculation of the pseudoscalar decay constants in
large $N_c$ chiral perturbation theory. The two angle scenario has been adopted
in a phenomenological analysis of the two-photon decay widths of the $\eta$ and
$\eta'$, the $\eta \gamma$ and $\eta' \gamma$ transition form factors,
radiative $J/\Psi$ decays, as well as of the  decay constants of the
pseudoscalar mesons \cite{FK1,FK2}.
The authors observe that within their phenomenological approach the assumption
of one mixing angle is not in agreement with experiment whereas the two-mixing
angle scheme leads to a very good description of the data.
As pointed out in these investigations the analysis with two different mixing
angles leads to a more coherent picture than the canonical treatment with a
single angle.
In particular, the calculation of the pseudoscalar decay constants within the
framework of large $N_c$ chiral perturbation theory
${\it requires}$ two different mixing angles \cite{L}.
(A similar investigation was performed in \cite{H-S1} but with a different
parameterization.)

In addition, it was shown in \cite{BB1} that even at leading order $\eta$-$\eta'$
mixing does not obey the usually assumed one-mixing-angle scheme, if large $N_c$
counting rules
are not employed. One purpose of this work is to critically re-examine the contributions
from $\eta$-$\eta'$ mixing to the two-photon decays.

For the decays $P \rightarrow \gamma \gamma^*$ contributions
from higher chiral orders will become increasingly more important for larger photon
virtualities. In this case one-loop contributions turn out to be divergent
and must be renormalized by appropriate counterterms \cite{BBC1, DW, BBC2}.
The remaining finite parts of the relevant counterterms can be estimated by 
vector-meson exchange contributions \cite{BBC2}. In \cite{BBC2} the vector-mesons were included explicitly
and integrating them out from the effective theory produced contact interactions.
An underlying assumption of this approach is that the masses of the vector-mesons
are much larger than the involved momenta. 

Alternatively, the chiral Lagrangian for the Goldstone
bosons is able to reproduce a number of meson resonances such as the
$f_0$(980) and the $\rho$(770), when combined with non-perturbative Bethe-Salpeter approaches
which are employed in such a way so that they ensure unitarity \cite{OO, OOP}.
Within these approaches effective coupled channel potentials are derived
from the chiral meson Lagrangian and iterated in Bethe-Salpeter equations (BSEs).
The BSE generates dynamically quasi-bound states of the mesons and 
accounts for the exchange of resonances without including them explicitly.
The usefulness of this approach lies in the fact that from a small set of
parameters a large variety of data can be explained and no additional assumptions need to be
made on the couplings of the resonances.
In \cite{BB2} the approach has been extended to include the $\eta'$ and
successfully applied to the hadronic decays of $\eta$ and $\eta'$ in \cite{BB3}.
The main purpose of the present work is to embed the coupled channel approach
in the two-photon decays $P \rightarrow \gamma^{(*)} \gamma^{(*)} $ and to investigate
the importance of vector-mesons for these decays from a different perspective.
This will also allow us to give predictions for the double Dalitz decays with 
two off-shell photons which have not yet been studied experimentally. It may further
clarify the question whether double vector-meson dominance holds, which is 
also an important issue for
the anomalous magnetic moment of the muon and kaon decays \cite{Bij2}.

This work is organized as follows. Section 2 introduces the notation and the tree level
contributions to the decays which arise from the WZW Lagrangian. One-loop contributions
and mixing effects are described and discussed in Sec.~3. 
The application of the coupled channel
analysis to the decays is outlined in Sec.~4 and the numerical results are presented
in Sec.~5. We conclude with a summary of our findings.

\section{Decays at tree level}\label{sec:tree}
The effective Lagrangian for the pseudoscalar meson nonet
($\pi, K, \eta_8, \eta_0$) reads up to second order in the derivative expansion
\cite{KL1,KL2,GL,BW}
\begin{eqnarray}  \label{eq:mes1}
\Lagr^{(0+2)} &=& - V_0
+V_1 \langle \cder_{\mu} U^{\dagger} \cder^{\mu}U \rangle
+V_2 \langle U^\dagger \chi+\chi^\dagger U\rangle
+i V_3 \langle U^\dagger \chi-\chi^\dagger U\rangle
\nonumber\\&&\mathord{}
+V_4 \langle U^\dagger\cder^{\mu} U \rangle \langle U^\dagger\cder_{\mu} U \rangle
\end{eqnarray}
where we have presented only the terms relevant
for the present work.
The unitary matrix $U$ is a $3 \times 3$ matrix containing the Goldstone boson
octet ($\pi^\pm,\pn, K, \eta_8$) and the $\eta_0$. 
Its dependence on $\pn, \eta_8$ and $\eta_0$ is given by
\begin{equation}
U=\exp\bigl(\mathrm{diag}(1,-1,0)\cdot i\pn/ \decay
+\mathrm{diag}(1,1,-2)\cdot i\eta_8/ \sqrt{3}\decay
+i\sqrt{2}\eta_0/\sqrt{3}\decay+\ldots\bigr).
\end{equation}
The expression $\langle \ldots \rangle$ denotes 
the trace in flavor space, $\decay$ is the pion decay constant in the chiral limit
and the quark mass matrix $\mathcal{M} = \mbox{diag}(m_u,m_d,m_s)$
enters in the combination  $\chi  = 2 B \mathcal{M} $
with $B = - \langle  0 | \bar{q} q | 0\rangle/ \decay^2$ being the order
parameter of the spontaneous symmetry violation.
The covariant derivative is defined by
\begin{eqnarray}
\cder_{\mu} U  &=&  \partial_{\mu} U - i ( v_{\mu} + \tilde{a}_{\mu}) U
                     + i U ( v_{\mu} - \tilde{a}_{\mu}) . 
\end{eqnarray}
The dependence of the effective Lagrangian on the running scale of QCD
due to the anomalous dimension of the singlet axial current
$A_\mu^0 = \frac{1}{2} \bar{q} \gamma_\mu \gamma_5 q$
is absorbed into the factor $\sqrt{\lambda}$
in the axial-vector connection $\tilde a_\mu$ 
\begin{equation}\label{eq:scalea0}
\tilde a_\mu=a_\mu + \frac{\sqrt{6\lambda}-f}{3f} \langle a_\mu \rangle.
\end{equation}
which is the scale independent combination of the octet and singlet parts of the
external axial-vector field $a_\mu$, cf. \cite{BW} for details.
Due to its scale dependence, $\sqrt{\lambda}$ cannot be determined from experiment, and
all quantities involving it are unphysical.

The coefficients $V_i$ are functions of 
$\eta_0$,
$V_i(\eta_0/\decay)$,
and can be expanded in terms of this variable. At a given order of
derivatives of the meson fields $U$ and insertions of the quark mass matrix 
$\mathcal{M}$ one obtains an infinite string of increasing powers of 
$\eta_0$ with couplings which are not fixed by chiral symmetry.
Parity conservation implies that the $V_i$ are all even functions
of $\eta_0$ except $V_3$, which is odd, and
$V_1(0) = V_2(0) = V_1(0)-3V_4(0)=\frac{1}{4}\decay^2$
 gives the correct  normalization
for the quadratic terms of the mesons.
The potentials $V_i$ are expanded in the singlet field $\eta_0$ 
\begin{eqnarray}\label{eq:vexpand}
V_i\Big[\frac{\eta_0}{\decay}\Big] &=& \coeffv{i}{0} + \coeffv{i}{2}
\frac{\eta_0^2}{\decay^2} +
\coeffv{i}{4} \frac{\eta_0^4}{\decay^4} + \ldots
\qquad \mbox{for} \quad i= 0,1,2,4 \nonumber \\
V_3\Big[\frac{\eta_0}{\decay}\Big] &=& \coeffv{3}{1} \frac{\eta_0}{\decay} + 
\coeffv{3}{3} \frac{\eta_0^3}{\decay^3}
+ \ldots \quad 
\end{eqnarray}
with expansion coefficients $\coeffv{i}{j}$ to be determined phenomenologically.

The Lagrangian $\Lagr^{(0+2)}$ contains only terms of natural parity.
The photonic decays $\pi \rightarrow \gamma \gamma, \eta \rightarrow \gamma \gamma, 
\eta' \rightarrow \gamma \gamma$, on the other hand, are not covered by $\Lagr^{(0+2)}$, since they arise
from the unnatural parity part of the effective Lagrangian
which collects the terms that are proportional to the tensor
$\epsilon_{\mu \nu \alpha \beta}$.  
Within the effective theory the chiral anomalies of the underlying QCD Lagrangian
are accounted for by the WZW term \cite{WZ, W, KL1} \footnote{Note that for our
purposes we can safely set the singlet axial vector field $\langle a_\mu \rangle$
and the derivative of the QCD vacuum angle, $\partial_\mu \theta$, to zero in $S_{\scriptscriptstyle{WZW}}$ 
which
enables us to work with the renormalization group invariant form of the anomaly.}
\beq  \label{eq:wzw}
S_{\scriptscriptstyle{WZW}} (U,v) = \int d^4 x  
\mathcal{L}_{\scriptscriptstyle{WZW}} = - \frac{i }{80 \pi^2} \int_{M_5} \langle \Sigma^5 \rangle
    -  \frac{i}{16 \pi^2} \int_{M_4} W(U,v)
\eeq
where
\beq  
W(U,v) = \langle \Sigma U^\dagger dv U v + \Sigma v dv + \Sigma dv v - i \Sigma^3 v  \rangle - 
         (U \leftrightarrow U^\dagger)
\eeq
with $\Sigma = U^\dagger dU$ and we set $N_c=3$ for the number of colors. 
We have displayed only the pieces of the Lagrangian relevant for the present work
and adopted the differential form notation of
\cite{KL1}
\beq
v = dx^\mu v_\mu , \qq d = dx^\mu \partial_\mu
\eeq
with the Grassmann variables $dx^{\mu}$ which yield the volume element $dx^\mu
dx^\nu dx^\alpha dx^\beta = \epsilon^{\mu \nu \alpha \beta} d^4x$.
The operation $(U \leftrightarrow U^\dagger)$ indicates the interchange of $U$ and 
$U^\dagger$. The first integral on the right-hand side
in Eq.~(\ref{eq:wzw}) is taken over the five-dimensional
manifold $M_5$ which is the direct product of the Minkowski space $M_4$ and
the finite interval $0\le x^5 \le 1$, while
the Grassmann algebra is supplemented by the fifth element $dx^5$. 
The matrix-valued fields $U(x,x^5)$ in the first integral
are functions on this five-dimensional manifold and interpolate smoothly
between the identity matrix $U(x,0)= \mathds{1}$ and the matrix $U(x)=U(x,1)$.
The value of the first integral is independent on the particular choice
of the interpolating functions $U(x,x^5)$, whereas the integration for the second
term in Eq.~(\ref{eq:wzw}) extends only over the Minkowski space $M_4$.
The external vector field $v= -e Q A$ describes the coupling of the photon field
$A=dx^\mu A_\mu$ to the mesons with $Q= \frac{1}{3} \mbox{diag}(2,-1,-1)$ being the charge matrix
of the light quarks.

At tree level only the terms quadratic in the vector fields $v$ contribute
and one arrives at the following pieces of the WZW Lagrangian
\beqa
d^4 x \mathcal{L}_{\scriptscriptstyle{WZW}} &=&  -  \frac{i}{16 \pi^2} \langle  U^\dagger dU U^\dagger dv U v  
         + U^\dagger dU v dv + U^\dagger dU dv v \no \\
  &&  \qq \q - U dU^\dagger U dv U^\dagger v  
         - U dU^\dagger v dv - U dU^\dagger dv v  \rangle  + \ldots \; .
\eeqa
However, this is not the whole story, since there exist further terms at
fourth chiral order which are gauge invariant and do not contribute to the
chiral anomalies. A complete set of these terms has been given in \cite{KL1},
out of which the following three contact terms contribute 
\beq \label{eq:ct4}
d^4 x \mathcal{L}_{\scriptscriptstyle{ct}}^{(4)} = W_1  \langle  U dv U^\dagger dv \rangle
  + W_2  \langle   dv  dv \rangle + i W_3 \langle dU dU^{\dagger} dv + dU^{\dagger} dU dv \rangle ,
\eeq
where the potentials $W_{1,2,3}$ are odd functions of the singlet field $\eta_0$.
The first two terms contribute already at tree level, whereas the last term
represents a vertex for a one-loop diagram and will be discussed in the next section.
Following the steps of \cite{BW} it is straightforward to see that
these terms are needed to absorb the QCD renormalization scale dependence via a set
of redefinitions of the potentials $W_i$ into
the factor $\sqrt{\lambda}$ of $\tilde a_\mu$ in Eq.~(\ref{eq:scalea0}).

\begin{figure}
\centering
\includegraphics[width=0.25\textwidth]{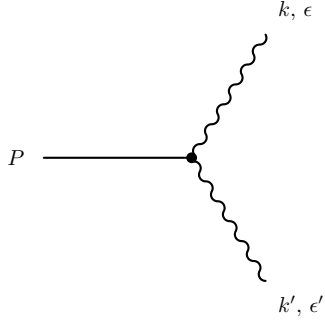}
\caption{Tree diagram of the decay $P \to \gamma^{(*)} \gamma^{(*)}$,
         where $k^{(\prime)}$ and $\epsilon^{(\prime)}$ denote the momenta and polarizations
         of the photons, respectively.}
\label{fig1}
\end{figure}

The decay at tree level is depicted in Figure~\ref{fig1}
and the pertinent
amplitudes  are given by
\beq  \label{eq:tree}
\mathcal{A}^{\textit{(tree)}} (P \rightarrow \gamma^{(*)} \gamma^{(*)}) = 
  e^2 k_\mu \epsilon_\nu k'_\alpha \epsilon'_\beta
  \epsilon^{\mu \nu \alpha \beta } \frac{1}{4 \pi^2 f} \alpha_P^{\textit{(tree)}}
\eeq
with
\beq
\alpha_{\pi^0}^{\textit{(tree)}} = 1 , \qq \qq \alpha_{\eta}^{\textit{(tree)}} = \frac{1}{\sqrt{3}} ,
 \qq \qq \alpha_{\eta'}^{\textit{(tree)}} = 2 \sqrt{\frac{2}{3}} - \frac{16 \pi^2}{3}
  ( w_1^{(1)} +  w_2^{(1)} ) ,
\eeq
where $w_1^{(1)}$, $w_2^{(1)}$ are the leading expansion coefficients
of the potentials $W_1$ and $W_2$, respectively. 
It is important to note that in our scheme $\eta$-$\eta'$ mixing is of second chiral order
\cite{BB1} and will modify the sub-leading order, {\it i.e.} the
one-loop order of the decay amplitudes.
This is in contradistinction to large $N_c$ ChPT where $\eta$-$\eta'$ mixing 
contributes at leading order.

\section{One-loop contributions}\label{sec:one-loop}
\begin{figure}
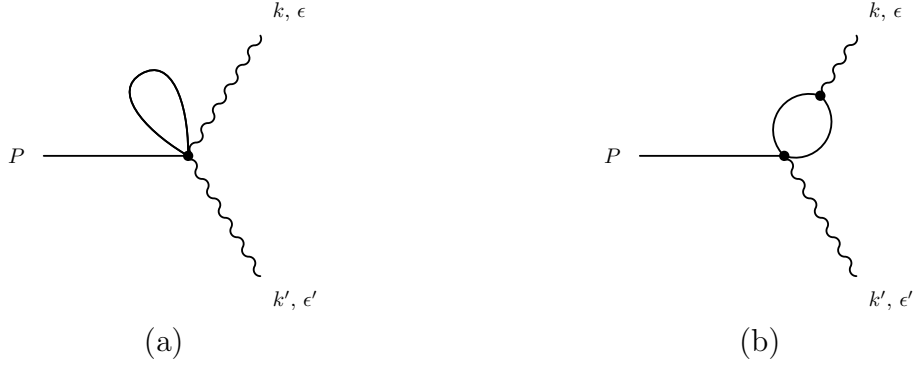

\centering
\begin{minipage}[b]{0.35\textwidth}
\centering
\includegraphics[scale=0.8]{feynps.2} \\
(a)
\end{minipage}
\hspace{0.1\textwidth}
\begin{minipage}[b]{0.35\textwidth}
\centering
\includegraphics[scale=0.8]{feynps.3} \\
(b)
\end{minipage}
\caption{One-loop diagrams contributing to $P \to \gamma^{(*)} \gamma^{(*)}$. 
         In (b) the crossed diagram is not shown.}
\label{fig2}
\end{figure}

The one-loop diagrams for $P \rightarrow \gamma^{(*)} \gamma^{(*)}$ 
of order $\mathcal{O}(p^6)$ have anomalous vertices from the
following pieces of the WZW Lagrangian
\beqa
d^4 x \mathcal{L}_{\scriptscriptstyle{WZW}} &=&  -  \frac{i}{16 \pi^2} \langle  U^\dagger dU U^\dagger dv U v  
        + U^\dagger dU v dv + U^\dagger dU dv v -i U^\dagger dU U^\dagger dU U^\dagger dU v \rangle \no \\
  & &   - (U \leftrightarrow U^\dagger) + \ldots \; .
\eeqa
After expanding $U= \exp ( \frac{i \sqrt{2}}{f} \phi)$ in the meson fields $\phi$
we arrive at the expression
\beq
d^4 x \mathcal{L}_{\scriptscriptstyle{WZW}} =
 -  \frac{\sqrt{2}}{8 \pi^2 f^3}  \Big( \langle  d\phi \, [ \phi, [\phi\, ,v\,  dv\, ]] \rangle
   -     \langle  d\phi\,  [ \phi \, , dv ] \, [\phi,\, v ] \rangle
    - 2 i \langle  d\phi \, d\phi \, d\phi \, v \rangle  \Big) + \ldots \; ,
\eeq
where we have only presented the relevant terms for the one-loop diagrams. 
The one-loop diagrams at order $\mathcal{O}(p^6)$ contributing to the decays
$P \rightarrow \gamma^{(*)} \gamma^{(*)} $ are shown in Fig.~\ref{fig2}.

The tadpole contributions in Fig.~\ref{fig2}a read
\beq   \label{eq:tad}
\mathcal{A}^{\textit{(tad)}} (P \rightarrow \gamma^{(*)} \gamma^{(*)}) = 
  e^2 k_\mu \epsilon_\nu k'_\alpha \epsilon'_\beta
  \epsilon^{\mu \nu \alpha \beta } \frac{1}{4 \pi^2 F_P^3} \Big(  
    \beta_P^{\textit{(tad)},\pi} \Delta_\pi   +  \beta_P^{\textit{(tad)},K} \Delta_K \Big)
\eeq
with $\Delta_\phi$ being the finite parts of the tadpoles
\beq
\Delta_\phi =  \left( \int \frac{d^d l}{(2 \pi)^d} \frac{i}{l^2 - m_\phi^2 +i\epsilon} \right)_{\textit{finite}}
 = \frac{m_\phi^2}{16 \pi^2} \,  \ln \frac{m_\phi^2}{\mu^2} ,
\eeq
and $\mu$ is the scale introduced in dimensional regularization.
In the present work we are only concerned with the finite pieces of the diagrams
and neglect the divergent portions throughout. 
The coefficients $\beta_P^{\textit{(tad)},\phi} $ are given by
\begin{equation}
\begin{array}{lcllcllcl}
\beta_\pi^{\textit{(tad)},\pi} & = & -\dfrac{4}{3} , \qq \qq 
& \beta_\eta^{\textit{(tad)},\pi} & = & -\dfrac{2}{\sqrt{3}} , \qq \qq   
& \beta_{\eta'}^{\textit{(tad)},\pi} & = & -2 \sqrt{\dfrac{2}{3}} + 16 \pi^2 w_{1}^{(1)} , \\
\beta_\pi^{\textit{(tad)},K} & = & -\dfrac{5}{3} ,  \qq \qq 
& \beta_\eta^{\textit{(tad)},K} & = & -\dfrac{1}{\sqrt{3}} , \qq \qq   
& \beta_{\eta'}^{\textit{(tad)},K} & = & -2 \sqrt{\dfrac{2}{3}} + 16 \pi^2 w_{1}^{(1)} .
\end{array}
\end{equation}
We have furthermore replaced the pion decay constant in the chiral limit, $f$,
by the expressions $F_P$ which include the one-loop corrections.
The corrections have been calculated in $U(3)$ ChPT without imposing
large $N_c$ counting rules and in infrared regularization \cite{BB1}. 
We can employ the
same formulae also for the present work by noting that only tadpoles contribute
at one-loop order to the decay constants. In infrared regularization the $\eta'$ tadpole
vanishes, whereas the tadpoles for the Goldstone boson octet remain unaltered.
This implies that the $\eta'$ tadpole does not contain any infrared physics and can be absorbed 
completely into the low-energy constants (LECs) of the effective Lagrangian.
It is neither a function of the Goldstone boson masses nor of the external momenta,
it is just a constant. We will therefore assume that $\eta'$ tadpole contributions have been
compensated by renormalizing the LECs appropriately and will work with the
renormalized values without indicating it explicitly.

The expansions of the decay constants in terms of the Goldstone boson masses
up to one-loop order are given here for completeness \cite{BB1}
\begin{eqnarray}\label{eq:axcouptwo}
F_\pi&=&\decay\Big[1+4 \cbeta{4}{0} \frac{2 m_K^2 + m_\pi^2}{ \decay^2}
+4 \cbeta{5}{0} \frac{m_\pi^2}{\decay^2}
-\frac{\tad{\pi}+\sfrac{1}{2}\tad{K}}{\decay^2}\Big],
\nonumber\\
F_{\eta}&=&\decay\Big[1+ 4 \cbeta{4}{0} \frac{2 m_K^2 + m_\pi^2}{\decay^2}
+4 \cbeta{5}{0} \frac{m_\eta^2}{\decay^2}
-\frac{\sfrac{3}{2}\tad{K}}{\decay^2}\Big],
\nonumber\\
F_{\eta'}&=& \frac{f}{\sqrt{6 \lambda}} F_{\eta' 0} = \decay \Big[1
+4 \frac{2 m_K^2 + m_\pi^2}{3 \decay^2}
 \big( 3 \cbeta{4}{0} + \cbeta{5}{0}  -9 \cbeta{17}{0} + 3 \cbeta{18}{0} 
\nonumber\\&&\qquad \qq \qq \qq \qq \qq \mathrel{}
+ 3 \cbeta{46}{0}  + 9 \cbeta{47}{0} - 3 \cbeta{53}{0}
  - \frac{3}{2} \sqrt{6} \cbeta{52}{1}    \big) \Big] .
\end{eqnarray}
The LECs $\cbeta{i}{j}$ originate from the natural parity part of the
effective Lagrangian at fourth chiral order 
\begin{equation}\label{eq:l4}
\Lagr^{(4)}=\tsum\nolimits_k \beta_k O_k,
\end{equation}
where the contributing fourth order operators are 
\begin{equation}  \label{eq:op}
\begin{array}{ll}
O_{\phantom{0}4}=-\langle C^\mu C_\mu\rangle\langle M\rangle, \qq \qq &
O_{\phantom{0}5}=-\langle C^\mu C_\mu M\rangle,\\
O_{17}=\langle C^\mu \rangle\langle C_\mu\rangle\langle M\rangle,&
O_{18}=-\langle C^\mu \rangle\langle C_\mu M\rangle,\\
O_{46}=2 i \langle \tilde a_\mu \rangle \langle C^\mu M\rangle,&
O_{47}=2 i \langle \tilde a_\mu \rangle \langle C^\mu \rangle \langle M\rangle,\\
O_{52}=-2 \langle M\rangle \partial^\mu \langle \tilde a_\mu \rangle ,&
O_{53}=2 i \langle N\rangle \partial^\mu \langle \tilde a_\mu \rangle ,
\end{array}
\end{equation}
and we made use of the following abbreviations 
\begin{equation} \label{eq:abr}
C_{\mu}=U^\dagger \cder_\mu U, \qq \qq 
M=U^\dagger \chi+\chi^\dagger U,\qq \qq 
N=U^\dagger \chi-\chi^\dagger U.
\end{equation}
The $\beta_k$ are functions of the singlet field, $\beta_k(\eta_0)$, and can be
expanded in the same manner as the $V_i (\eta_0)$.
Note that the $\beta_k$ also have divergent pieces, in order to compensate the
divergences from the loops, but in the present work we are only concerned
with the finite parts and use the same notation for simplicity.
The decay constant of the $\eta'$ related to the singlet axial-vector current,
$F_{\eta' 0}$, is defined by the matrix element
\beq
\langle 0| \sfrac{1}{\sqrt{6}} \bar{q} \gamma_\mu \gamma_5  q |\eta'\rangle=i p_\mu F_{\eta' 0} .
\eeq
Due to the anomalous dimension of the singlet axial
current, $F_{\eta' 0}$ depends on the running scale of QCD and
its value cannot be determined phenomenologically. 
In order to account for the chiral corrections
for $f$, we have therefore employed the scale invariant ratio 
$F_{\eta'}=\frac{f}{\sqrt{6 \lambda}} F_{\eta' 0}$ in the expression for the $\eta'$
decay amplitude in Eq.~(\ref{eq:tad}).

We replace the pion decay constant in the chiral limit, $f$, by the chirally 
corrected decay constants $F_P$ also
in the tree level result, Eq.~(\ref{eq:tree}). 
This replacement yields corrections at $\mathcal{O}(p^6)$ which must be compensated.
If the decay constants $F_P$ in Eq.~(\ref{eq:axcouptwo}) are written as $F_P = f(1+ \delta F_P/f^2)$,
it induces at sixth chiral order the corrections
\beq  \label{eq:deccor}
\mathcal{A}^{(f)} (P \rightarrow \gamma^{(*)} \gamma^{(*)}) = 
  e^2 k_\mu \epsilon_\nu k'_\alpha \epsilon'_\beta
  \epsilon^{\mu \nu \alpha \beta } \frac{1}{4 \pi^2 F_P^3} \alpha_P^{\textit{(tree)}} \delta F_P .
\eeq

The contributions from diagram \ref{fig2}b yield
\beqa  \label{eq:1-loop}
\mathcal{A}^{\textit{(uni)}} (P \rightarrow \gamma^{(*)} \gamma^{(*)}) &=& 
  e^2 k_\mu \epsilon_\nu k'_\alpha \epsilon'_\beta
  \epsilon^{\mu \nu \alpha \beta } \frac{1}{4 \pi^2 F_P^3} \Big(
    \beta_P^{\textit{(uni)},\pi} [ I(m_\pi^2;k^2) + I(m_\pi^2;k'^2) ]  \no \\
  &&  \qq \qq \qq \qq
    +  \beta_P^{\textit{(uni)},K} [ I(m_K^2;k^2) + I(m_K^2;k'^2) ]\Big) .
\eeqa
The integral $I$ is given by
\beq \label{eq:int1}
I(m^2;p^2) =   \frac{2 }{3} \Big(  \frac{1}{2} \Delta  +  (m^2 - \frac{p^2}{4}) G_{mm}(p^2)   
      + \frac{1}{96 \pi^2} (p^2 -6 m^2)     \Big) ,
\eeq
where $G$ is the finite part of the scalar one-loop integral
\begin{equation}
G_{m \bar{m}}(p^2) =\mathrel{}
\left( \int\frac{\,d^d l}{(2\pi)^d}\,\frac{i}{(l^2-m^2+i \epsilon)( (l-p)^2-\bar m^2+i \epsilon)} \right)_{\textit{finite}}
\end{equation}
which reads
{\arraycolsep0pt\begin{eqnarray}  \label{eq:g}
G_{m \bar{m}}(p^2) =&&  \mathrel{}
\frac{1}{16\pi^2}\bigg[-1+
\ln\frac{m \bar{m}}{\mu^2}
  +\frac{m^2-\bar{m}^2}{p^2}\ln\frac{m}{\bar{m}}
\nonumber\\&&\qquad\qquad\quad\mathord{}
  -\frac{2\sqrt{\deltaph_{m\bar{m}}(p^2)}}{p^2}\artanh\frac{
  \sqrt{\deltaph_{m\bar{m}}(p^2)}}{(m+\bar{m})^2-p^2}\bigg] ,
\nonumber\\
\deltaph_{m\bar{m}}(p^2)=&&\mathrel{}\big((m-\bar{m})^2-p^2\big)\big((m+\bar{m})^2-p^2\big).
\end{eqnarray}}%
For the particular case with on-shell photons the integral $I$
reduces to the simple expression $I(m^2;0)= \Delta$.
The coefficients $\beta_P^{\textit{(uni)},\phi} $ are given by
\begin{equation}
\begin{array}{lcllcllcl}
\beta_\pi^{\textit{(uni)},\pi} & = & 1 , \qq \qq &  \beta_\eta^{\textit{(uni)},\pi} & = & \dfrac{1}{\sqrt{3}} , 
 \qq \qq   & \beta_{\eta'}^{\textit{(uni)},\pi} & = & \sqrt{\dfrac{2}{3}} - 16 \pi^2 w_{3}^{(1)} , \\
\beta_\pi^{\textit{(uni)},K} & = & 1 ,  \qq \qq &  \beta_\eta^{\textit{(uni)},K} & = & \dfrac{1}{\sqrt{3}} , 
 \qq \qq   & \beta_{\eta'}^{\textit{(uni)},K} & = & \sqrt{\dfrac{2}{3}} - 16 \pi^2 w_{3}^{(1)} .
\end{array}
\end{equation}
%

\subsection{Wave-function renormalization,
$\mbox{\boldmath$\eta$}$-$\mbox{\boldmath$\eta'$}$ mixing, and counterterms}\label{sec:wave}
There are further chiral corrections at $\mathcal{O}(p^6)$ which arise from
wave-function renormalization and $\eta$-$\eta'$ mixing. 
The relation between the original $SU(3)$ fields ($\pn,\eta_8, \eta_0$)
and the renormalized states ($\pi^0,\eta, \eta'$) 
without employing large $N_c$ counting rules has already been derived in \cite{BB1}.
It is given in terms of the $3 \times 3$ matrix $\mathds{1} + R^{(2)}$,
$(\pn,\eta_8,\eta_0)^T=(\mathds{1} + R^{(2)})(\pi^0,\eta,\eta')^T$, with
\begin{eqnarray} \label{eq:trafo}
R^{(2)}_{\pn \pi^0}&=&\bigl(-4 [2m_K^2 + m_\pi^2] \cbeta{4}{0}
-4m^2_{\pi}\cbeta{5}{0}
+\sfrac{1}{3}\tad{\pi}+\sfrac{1}{6}\tad{K}\bigr)/F_\pi^2,
\nonumber\\
R^{(2)}_{8\eta}&=&\bigl(-4 [2m_K^2 + m_\pi^2] \cbeta{4}{0}-4m_{\eta}^2\cbeta{5}{0}
+\sfrac{1}{2}\tad{K}\bigr)/F_\eta^2,
\nonumber\\
R^{(2)}_{8\eta'}&=&\frac{8\sqrt{2}}{3} (m_K^2 - m_\pi^2 ) 
\bigl(2\mtop \beta_{5,18} - \cvtwid{2}{1}\bigr)\big/F_{\eta'}^2\mtop,
\nonumber\\
R^{(2)}_{0\eta}&=&\frac{8\sqrt{2}}{3} (m_K^2 - m_\pi^2 ) \cvtwid{2}{1}\big/F_\eta^2\mtop,
\nonumber\\
R^{(2)}_{0 \eta'}&=&-\frac{4}{3} (2m_K^2 + m_\pi^2) (3\cbeta{4}{0}+\cbeta{5}{0}-9\cbeta{17}{0}+3\cbeta{18}{0})/F_{\eta'}^2,
\end{eqnarray}
and all remaining entries vanish
\footnote{Note that 
we work in the isospin limit so that
the $\pn$ field does not undergo mixing with the $\eta_8$-$\eta_0$ system.}.
In Eq.~(\ref{eq:trafo}) we have used the abbreviations
\begin{eqnarray}\label{eq:abbrev}
\mtop&=&\frac{2\coeffv{0}{2}}{\decay^2},
\nonumber\\
\cvtwid{2}{1}&=&\sfrac{1}{4}\decay^2-\sfrac{1}{2}\sqrt{6}\coeffv{3}{1},
\nonumber\\
\beta_{5,18}&=&\cbeta{5}{0}+\sfrac{3}{2}\cbeta{18}{0}.
\end{eqnarray}
Within this scenario, $\eta$-$\eta'$ mixing contributes at next-to-leading order
which is in contradistinction to large $N_c$ ChPT \cite{GBH}, where it is a
leading order effect.
Inserting these relations into the tree level result Eq.~(\ref{eq:tree})
one obtains the following corrections at sixth chiral order
\beq  \label{eq:wfcor}
\mathcal{A}^{(Z)} (P \rightarrow \gamma^{(*)} \gamma^{(*)}) = 
  e^2 k_\mu \epsilon_\nu k'_\alpha \epsilon'_\beta
  \epsilon^{\mu \nu \alpha \beta } \frac{1}{4 \pi^2 F_P} \beta^{(Z)}_P 
\eeq
with the coefficients
\beqa
\beta^{(Z)}_{\pi^0} &=& \alpha_{\pi^0}^{\textit{(tree)}} R^{(2)}_{\pn \pi^0} \no \\
\beta^{(Z)}_{\eta} &=& \alpha_{\eta}^{\textit{(tree)}} R^{(2)}_{8 \eta}
       + \alpha_{\eta'}^{\textit{(tree)}} R^{(2)}_{0 \eta}   \no \\
\beta^{(Z)}_{\eta'} &=& \alpha_{\eta}^{\textit{(tree)}} R^{(2)}_{8 \eta'}
       + \alpha_{\eta'}^{\textit{(tree)}} R^{(2)}_{0 \eta'}  .
\eeqa
Summing all the contributions at order $\mathcal{O}(p^6)$ we verify that
for the $\pi^0$- and $\eta$-decays into two on-shell photons the dependence on the
regularization scale $\mu$ cancels out, {\it i.e.} the amplitudes are
finite and do not need to be renormalized by counterterms of the unnatural
parity part of the $p^6$ Lagrangian \cite{DHL, BBC1, DW, BBC2}.
For the process $\eta' \to \gamma \gamma$, to the contrary, the $w_{3}^{(1)}$ term of the unnatural  parity
Lagrangian at fourth chiral order induces a non-vanishing $\mu$ dependence via a loop. 
Also, for the decays with one or both photons off-shell the divergent parts do not cancel for any 
of the decaying mesons and again a residual scale dependence remains
which can be compensated by introducing counterterms at sixth order.
The explicit renormalization of the amplitudes which are induced by the WZW term
has been accomplished in \cite{DW, BBC2}
and is beyond the scope of the present investigation, where we restrict ourselves to the finite pieces
of the integrals and counterterms.

In the $SU(3)$ framework without an explicit $\eta'$, the Lagrangian of unnatural 
parity at $\mathcal{O}(p^6)$ has been constructed in \cite{EFS}.
The relevant terms contributing via tree diagrams to the decays are presented
in App. \ref{app:a}, where we also show an additional set of counterterms
which arises due to the extension to the $U(3)$ framework.
These terms can be reduced to the following structures
(restricting ourselves to the finite pieces of the coupling constants)
\begin{equation} \label{lagrct}
\begin{array}{lcclccl}
\mathcal{L}_{ct}^{(6)} & = &
- & \bar{w}_{1}^{(0)} \ \frac{16 \sqrt{2}}{f} \ \langle \phi \ d v \ \square \, d v \rangle
& + & \bar{w}_{2}^{(1)} \  \frac{16}{f} \ \eta_0 \ \langle d v \ \square \, d v \rangle \\
& & + & \bar{w}_{3}^{(0)} \ \frac{32 \sqrt{2}}{f} \langle \phi \,\chi \, dv \, dv \rangle
& + & \bar{w}_{4}^{(0)}\ \frac{32 \sqrt{2}}{f} \langle \phi \,\chi \rangle \langle dv \, dv \rangle \\
& & + & \bar{w}_{5}^{(1)} \, \ \frac{32}{f} \ \eta_0 \langle \chi \, dv \, dv \rangle
& + & \bar{w}_{6}^{(1)} \, \ \frac{32}{f} \ \eta_0 \langle \chi \rangle \langle dv \, dv \rangle ,
\end{array}
\end{equation}
where we employed the LECs of the original counterterms in App. \ref{app:a}.
Note that the first two terms only contribute to the decays with off-shell photons.
The contribution of $\mathcal{L}_{ct}^{(6)}$ to the decays reads
\beq \label{ctcontr}
\mathcal{A}^{\textit{(ct)}} =
(P \rightarrow \gamma^{(*)} \gamma^{(*)}) = 
  e^2 k_\mu \epsilon_\nu k'_\alpha \epsilon'_\beta
  \epsilon^{\mu \nu \alpha \beta } \frac{1}{4 \pi^2 F_P} \beta^{(ct)}_P
\eeq
with the coefficients
\beqa  \label{eq:ctcoeff}
\beta^{(ct)}_{\pi^0} & = & - \frac{64}{3} \pi^2 [(k^2 + {k'}^2) \bar{w}_{1}^{(0)} 
+ 4 m_{\pi}^2 \bar{w}_{3}^{(0)}] ,  \no \\
\beta^{(ct)}_{\eta} & = & - \frac{64}{3 \sqrt{3}} \pi^2 \left[
(k^2 + {k'}^2) \bar{w}_{1}^{(0)} - \frac{4}{3}(4 m_{K}^2 - 7 m_{\pi}^2) \bar{w}_{3}^{(0)}
- 32(m_{K}^2 - m_{\pi}^2) \bar{w}_{4}^{(0)} \right] , \no \\
\beta^{(ct)}_{\eta'} & = & - \frac{128}{3} \pi^2 \left[
(k^2 + {k'}^2)(\sqrt{\frac{2}{3}} \bar{w}_{1}^{(0)} - \bar{w}_{2}^{(1)})
+ \frac{4}{3}(m_{K}^2 + 2 m_{\pi}^2)(\sqrt{\frac{2}{3}} \bar{w}_{3}^{(0)} + \bar{w}_{5}^{(1)})
\right. \no \\
& & \qq \qq \left.
+ 4(2 m_{K}^2 + m_{\pi}^2)(\sqrt{\frac{2}{3}} \bar{w}_{4}^{(0)} + \bar{w}_{6}^{(1)})
\right]  .
\eeqa
%

\subsection{Numerical results at one-loop order}\label{sec:numloop}
Before implementing the unitarity corrections beyond one-loop within the Bethe Salpeter
approach, we would like to extract numerical results from the one-loop expressions
for the decays into two physical photons, $P \rightarrow \gamma \gamma$.
Combining all the contributions of the preceding sections 
we arrive at
\beq
\mathcal{A}^{\textit{(1-loop)}} = \mathcal{A}^{\textit{(tree)}} + \mathcal{A}^{\textit{(tad)}} + \mathcal{A}^{\textit{(uni)}}
+ \mathcal{A}^{(f)} + \mathcal{A}^{(Z)} + \mathcal{A}^{(ct)}  .
\eeq
The decay width $\Gamma$ is given by
\beq
\Gamma (P \rightarrow \gamma \gamma) = \frac{\alpha^2 m_P^3}{64 \pi^3 F_P^2} \; 
    |\,\beta_P^{\textit{(1-loop)}} \, |^{\,2}
\eeq
with $\alpha= e^2/4 \pi$ and
\beq
\beta_P^{\textit{(1-loop)}} = \alpha_P^{\textit{(tree)}} \left(1 + \frac{\delta F_P}{F_{P}^2} \right)
  + (\beta_P^{\textit{(tad)},\pi} + 2 \beta_P^{\textit{(uni)},\pi}) \frac{\Delta_\pi}{F_{P}^2}
  + ( \beta_P^{\textit{(tad)},K} + 2 \beta_P^{\textit{(uni)},K} )  \frac{\Delta_K}{F_{P}^2}
  + \beta_P^{(Z)} + \beta_P^{(ct)} .
\eeq
For comparison with previous work \cite{DHL, BBC1} we also present the explicit form of the 
$\beta_P^{\textit{(1-loop)}}$ coefficients
\begin{eqnarray}
\beta_{\pi^0}^{\textit{(1-loop)}} & = & 1 - \frac{256 \pi^2}{3} m_{\pi}^2 \bar{w}_{3}^{(0)} , \no \\
\beta_{\eta}^{\textit{(1-loop)}} & = & \frac{1}{\sqrt{3}} \left\{1 + \frac{1}{F_{\eta}^2} \frac{32}{3} 
  \left[1 - \sqrt{\frac{2}{3}} \ 4 \pi^2 (w_{1}^{(1)} + w_{2}^{(1)}) \right]
        (m_{K}^2 - m_{\pi}^2) \frac{\cvtwid{2}{1}}{\mtop} \right. \no \\
& & \qq \left. + \ \frac{256}{9} \pi^2 \ [(4 m_{K}^2 - 7 m_{\pi}^2) \bar{w}_{3}^{(0)} 
        + 24 (m_{K}^2 - m_{\pi}^2) \bar{w}_{4}^{(0)} ] \right\} , \no \\
\beta_{\eta'}^{\textit{(1-loop)}} & = & \left[2 \sqrt\frac{2}{3} - \frac{16 \pi^2}{3}
  (w_{1}^{(1)} + w_{2}^{(1)}) \right]
  \left\{1 + \frac{1}{F_{\eta'}^2} \left[ 4 (2 m_{K}^2 + m_{\pi}^2)
  (\cbeta{46}{0} + 3\cbeta{47}{0} - \cbeta{53}{0} - \sqrt{\frac{3}{2}} \cbeta{52}{1})
  \right] \right\} \no \\
& & \qq + \ \frac{1}{F_{\eta'}^2} \left[16 \pi^2 (w_{1}^{(1)} - 2w_{3}^{(1)})(\Delta_\pi + \Delta_K) 
        + \sqrt{\frac{2}{3}} \ \frac{8}{3} (m_{K}^2 - m_{\pi}^2) (2 \beta_{5,18}
        - \frac{\cvtwid{2}{1}}{\mtop}) \right] \no \\
& & \qq - \ \frac{512}{9} \pi^2 \left[ (m_{K}^2 + 2 m_{\pi}^2) (\sqrt{\frac{2}{3}} \bar{w}_{3}^{(0)} 
        + \bar{w}_{5}^{(1)}) + 3 (2 m_{K}^2 + m_{\pi}^2) (\sqrt{\frac{2}{3}} \bar{w}_{4}^{(0)}
        + \bar{w}_{6}^{(1)}) \right] .
\end{eqnarray}
For the pion decay constant we take the value $F_{\pi^0} = F_{\pi^+} = 92.4$~MeV,
while for $F_\eta$ we employ $F_\eta \approx 1.3 F_\pi$ which follows from
an analysis within the framework of conventional $SU(3)$ ChPT
that is similar to our expression at the order we are working \cite{GL}.
Since both the values of the contact terms $w_1^{(1)}$, $w_2^{(1)}$ and $w_3^{(1)}$,
which contribute to the $\eta'$ decay as well as 
to the $\eta$ decay due to $\eta$-$\eta'$ mixing, 
and the values of the $\mathcal{O}(p^6)$ couplings $\bar{w}_{1}^{(0)}, \dots, \bar{w}_{4}^{(0)},
\bar{w}_{5}^{(1)}, \bar{w}_{6}^{(1)}$
are unknown, we first set them by hand to zero in order to see,
if a reasonable fit to all three decay widths is possible without them.
The difference between both mixing angles for $\eta$-$\eta'$ mixing is proportional
to the parameter combination $ \cbeta{5}{0} + \frac{3}{2} \cbeta{18}{0} $ \cite{BB1}
which is dominated by $ \cbeta{5}{0}$, since $\cbeta{18}{0}$ represents an OZI 
violating correction. We will thus neglect $\cbeta{18}{0}$ and borrow the value for 
$\cbeta{5}{0} $ from conventional $SU(3)$ ChPT, $\cbeta{5}{0} = 1.4 \cdot 10^{-3}$
\cite{BEG}.
Furthermore, the parameter combination
$\cbeta{46}{0} + 3\cbeta{47}{0} - \cbeta{53}{0} - \sqrt{\frac{3}{2}} \cbeta{52}{1}$
represents a $1/N_c$ suppressed contribution to $F_{\eta'}$ in Eq.~(\ref{eq:axcouptwo})
with respect to $\cbeta{4}{0} + \frac{1}{3}
\cbeta{5}{0} - 3 \cbeta{17}{0} + \cbeta{18}{0}$. The value of the latter
combination has been estimated in \cite{BB1} to be roughly $0.15 \times 10^{-3}$
and therefore $\cbeta{46}{0} + 3\cbeta{47}{0} - \cbeta{53}{0} - \sqrt{\frac{3}{2}} \cbeta{52}{1}$
is expected to yield a tiny correction to $\beta_{\eta'}^{\textit{(1-loop)}}$,
so that for our purposes it can be set to zero.
The only remaining parameters are then $F_{\eta'}$ and $\cvtwid{2}{1}$
which we fit to the decay widths $\Gamma_{\pi^0} = 7.74$~eV, $\Gamma_\eta = 0.46$~keV,
and $\Gamma_{\eta'} = 4.28$~keV. 
These are the central experimental numbers
quoted by the Particle Data Group \cite{pdg}.
The fit yields $F_{\eta'} = 1.29 F_\pi$ and $\cvtwid{2}{1} = 1.15 F_\pi^2/4$.
The value for $\cvtwid{2}{1}$ is in agreement with the one derived
from an analysis of the pseudoscalar meson masses and decay constants \cite{BB1}.
The extracted value for $F_{\eta'} $, on the other hand, is slightly larger than
the values deduced within the one-mixing angle scheme, $F_{\eta'} \approx 1.1 F_\pi$ \cite{DHL, BBC1}.
This would indicate that in the present approach some of the OZI violating corrections for $F_{\eta'}$ in
Eq.~(\ref{eq:axcouptwo}) are comparable in size with the leading contribution $\cbeta{5}{0}$.
If, on the other hand, one fixes $F_{\eta'}\approx 1.1 F_\pi$, a fit with only $\cvtwid{2}{1}$ as a free
parameter is no longer possible even within the experimental error ranges.

We also compare our results with the one-mixing angle scheme by putting
$ \cbeta{5}{0}$ to zero. By fitting $F_{\eta'}$ and $\cvtwid{2}{1}$
to the decay widths we obtain $F_{\eta'} = 1.15 F_\pi$ and $\cvtwid{2}{1} = 1.15 F_\pi^2/4$
which would correspond to a mixing angle of about $-20^\circ$ in nice agreement
with  \cite{DHL, BBC1}.
If we include the contact terms $w_1^{(1)}$, $w_2^{(1)}$ and $w_3^{(1)}$, we have the freedom
to choose $F_{\eta'} = 1.1 F_\pi$ while still being able to match the experimental data. 
Changing the LECs $w_1^{(1)}$, $w_2^{(1)}$ and $w_3^{(1)}$ within a reasonable range of 
$-3.0 \times 10^{-3} \dots 3.0 \times 10^{-3}$ that is motivated by large $N_c$ considerations 
and comparison with the coefficients of the WZW term
we find a variation of the mixing parameter of $\cvtwid{2}{1} = (1.25 \pm 0.18) F_\pi^2/4$ which 
is commensurable with a mixing angle of $-21.3^\circ \pm 3.1^\circ$ in the one-mixing angle scheme. 
The lack of knowledge of the exact values of $w_1^{(1)}$, $w_2^{(1)}$ and $w_3^{(1)}$ 
induces a 15~\% uncertainty in the determination of the $\eta$-$\eta'$-mixing angle.

Since with the chosen parameters we are able to accommodate the experimental decay widths,
while being in agreement with results from the conventional $SU(3)$ framework,
there is no indication that the unknown parameters of sixth chiral order 
$\bar{w}_{1}^{(0)}, \dots, \bar{w}_{4}^{(0)}, \bar{w}_{5}^{(1)}, \bar{w}_{6}^{(1)}$ 
should have values significantly different from zero which would lead to sizeable 
contributions for the decay widths.
Within the one-loop calculation presented here, we can thus neglect them.

Note also that we work in the isospin limit of equal up- and down-quark masses. The effect of isospin 
breaking is only important for the decay of the $\pi^0$. It has been discussed in 
\cite{GBH, M, AM} and yields a correction of about 5~\%
to the $\Gamma_{\pi^0}$ decay width.

\section{Unitarity corrections beyond one-loop}\label{sec:unitarity}
In the preceding sections, we have identified the different types of contributions
which arise from both the WZW and the unnatural parity $p^4$ Lagrangian at tree level 
and next-to-leading order, while
the numerous counterterms of the unnatural parity $p^6$ Lagrangian were neglected.
The proliferation of such counterterms makes a unique fit to data impossible \cite{BBC1, DW}
and one must resort to model-dependent assumptions.
One possible way of estimating the size of the different parameters is to calculate
contributions from vector-meson exchange \cite{BBC2}, {\it e.g.}, by employing
the hidden symmetry formulation of Bando et al. \cite{BKY, Fuj}.
To this end, an effective Lagrangian of unnatural parity and including
the nonet of the lowest-lying vector-mesons, $V$, is constructed. Then the vector-mesons are integrated
out of the effective theory under the assumption that their masses are much larger
than the momenta which generates counterterms of order $p^6$ and higher.
The unknown counterterms of the unnatural parity $p^6$ Lagrangian without
vector-mesons can thus be written in terms of a few parameters of the vector-meson Lagrangian.
The coupling constants of the latter can be extracted up to a sign and within
certain error bars from radiative decay widths of the vector-mesons,
such as $\omega \rightarrow \pi^0 \gamma$ and $\rho^+ \rightarrow \pi^+ \gamma$.
(More recently, the parameters of the vector-meson Lagrangian have been
constrained by the short-distance behavior of QCD \cite{RPP}.)
This approach has been applied to the decay process $P \rightarrow \gamma \gamma^*$ in \cite{BBC2}
which amounts to the two-step chain $P \rightarrow V V \rightarrow \gamma \gamma^*$
with the virtual vector-mesons being treated as infinitely heavy states.
Besides giving estimates for the LECs of the higher order $p^6$ Lagrangian this procedure reproduces also
the experimental $P \rightarrow \gamma \gamma^*$ slopes \cite{BBC2} and must be added to
the one-loop contributions to the slopes which are extracted from Eq.~(\ref{eq:1-loop}). 
These are much smaller in magnitude and far away from the experimental data \cite{Dzh, Aih, Beh}.
We will discuss this point in more detail in Sec.~\ref{sec:results}
when we present the numerical results.

On the other hand, some of the vector-mesons such as the $\rho(770)$
 can be described as bound states of two Goldstone bosons
\cite{OOP}. Effective potentials for meson-meson scattering
are derived from the chiral Lagrangian for
the pseudoscalar mesons and iterated within a coupled channel BSE which satisfies
unitarity constraints for the partial-wave amplitudes. With a small set
of parameters a large variety of meson-meson scattering data could be explained up to center-of-mass
energies of 1.2 GeV.\@
In \cite{BB2} 
it was shown that the inclusion of the $\eta'$ in this framework does not spoil
the results from \cite{OOP} for energies below 1.2 GeV as one would expect naively.
We  adopt the approach of \cite{BB2} here and will not include the vector-mesons
explicitly in the theory. 
They will be rather generated  from composed states of two pseudoscalar mesons, whereas
composed states of three pseudoscalar mesons such as the $\omega(782)$ are beyond
the scope of the present investigation and will be neglected.
The additional parameters in the coupled channel
analysis are fixed from a fit to the $p$-wave phase shifts and we do not
have to deal with additional coupling constants for the vector-mesons
which always introduce a theoretical uncertainty.
Furthermore, our approach is suited to go to higher photon virtualities
which presents a problem in the resonance saturation scheme, since
large momenta in the vector-meson propagators cannot be simply dropped
when integrating them out of the effective theory.

Finally, our investigation includes and distinguishes the processes 
$P \rightarrow V \gamma^{(*)} \rightarrow \gamma^{(*)} \gamma^{(*)}$
and $P \rightarrow V V \rightarrow \gamma^{(*)} \gamma^{(*)}$,
where the photons can be either on- or off-shell.
In the approach with explicit vector-mesons, which has been applied so far only
to the decays with at least one on-shell photon, the first decay chain is not directly present.
It is rather derived from the latter one by integrating out the vector-meson coupled to the
physical photon.
Therefore, for the decays with one on-shell photon no clear distinction 
can be made in this scenario between a decay chain with
one or two virtual vector-mesons as suggested by complete Vector-Meson Dominance.
A measurement of the decays with two off-shell photons would help to clarify the situation.

\subsection{Bethe-Salpeter equation}\label{sec:bse}
\begin{figure}
\centering
\includegraphics[width=0.3\textwidth]{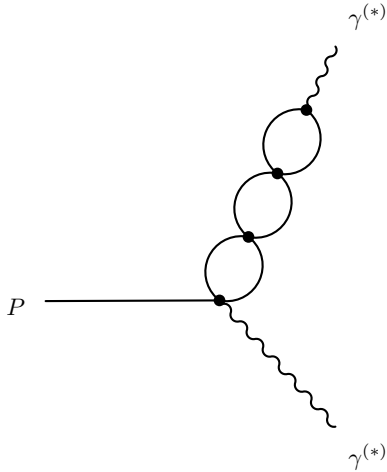}
\caption{Sample rescattering process in the decay $P \to \gamma^{(*)} \gamma^{(*)}$.}
\label{fig3}
\end{figure}

The underlying idea of our approach is as follows. The incoming pseudoscalar meson $P$
can decay via a vertex of either the WZW Lagrangian, $\mathcal{L}_{\scriptscriptstyle{WZW}}$,
or the unnatural parity Lagrangian at fourth chiral order, 
$\mathcal{L}_{\scriptscriptstyle{ CT}}^{(4)}$ in Eq.~(\ref{eq:ct4}), directly
into one of the following three channels: two photons, a photon and two pseudoscalar mesons, 
or four pseudoscalar mesons.
The mesons can combine to pairs and rescatter an arbitrary number of times
before they eventually couple to a photon,
see Fig.~\ref{fig3} for illustration.

The rescattering process can be described by application of the BSE
which generates the propagator for two interacting particles.
A similar approach has already been successfully employed for 
the hadronic decay modes of $\eta$ and $\eta'$
\cite{BB3}, where only $s$-wave amplitudes were considered.
Here, we extend it to photonic decays and $p$-waves.
In this section, we describe the BSE approach for meson-meson scattering in general focusing on the
$p$-wave contributions, while in the next section this method will be embedded into the
two-photon decays of $\pi^0, \eta$ and $\eta'$.

In order to describe meson-meson scattering accurately within the coupled channel approaches,
it is necessary to construct the interaction kernel for the two mesons from the effective
Lagrangian up to fourth chiral order. In addition to $\Lagr^{(0+2)}$
and the Operators $O_4,O_5,O_{17}, O_{18}$ from $\Lagr^{(4)}$
in Eq.~(\ref{eq:op}), we also include at fourth chiral order the terms
\begin{equation}
\begin{array}{ll}
O_{0\phantom{0}}=\langle C^\mu C^\nu C_\mu C_\nu\rangle,&
O_{1\phantom{0}}=\langle C^\mu C_\mu\rangle\langle C^\nu C_\nu\rangle,\\
O_{2\phantom{0}}=\langle C^\mu C^\nu\rangle\langle C_\mu C_\nu\rangle,&
O_{3\phantom{0}}=\langle C^\mu C_\mu C^\nu C_\nu\rangle,\\
O_{13}=-\langle C^\mu\rangle\langle C_\mu C^\nu C_\nu\rangle,&
O_{14}=-\langle C^\mu\rangle\langle C_\mu\rangle \langle C^\nu C_\nu\rangle,\\
O_{15}=-\langle C^\mu\rangle\langle C^\nu\rangle \langle C_\mu C_\nu\rangle,\qquad&
O_{16}=\langle C^\mu\rangle\langle C_\mu\rangle\langle C^\nu\rangle\langle C_\nu\rangle,\\
O_{21}=\langle C^\mu C_\mu i N\rangle,&
O_{22}=\langle C^\mu C_\mu\rangle\langle i N\rangle,\\
O_{23}=\langle C^\mu \rangle\langle C_\mu i N\rangle,&
O_{24}=\langle C^\mu \rangle\langle C_\mu\rangle\langle i N\rangle .\\
\end{array}
\end{equation}
Usually the $\beta_0$ term is not presented in conventional ChPT, since
there is a Cayley-Hamilton matrix identity that enables one to
remove this term leading to modified coefficients
$\beta_i$, $i=1,2,3,13,14,15,16$ \cite{H-S1},
but for our purposes
it turns out to be more convenient to include it.
Hence, we do not make use of the Cayley-Hamilton identity
and keep all couplings, in order to work with the most general expressions
in terms of these parameters. One can then drop one of the
$\beta_i$ involved in the Cayley-Hamilton identity
at any stage of the calculation.

From the contact interactions of the effective Lagrangian up to fourth chiral 
order we derive
the center-of-mass scattering amplitude $A(\theta)$, where $\theta$
is the c.\,m.\ scattering angle.
The partial-wave expansion for $A(\theta)$
reads
\begin{equation}
A(\theta) = \sum_{l=0}^2 (2l+1) A_l P_l(\cos \theta),
\end{equation}
where $A_l$ is the effective potential for angular momentum $l$ obtained
from the contact interactions and $P_l$
is the $l^{\textrm{th}}$ Legendre polynomial.
It is most convenient to work in the isospin basis and to characterize the
meson-meson states by their total isospin.
For the $p$-waves, {\it e.g.}, the relevant two-particle states have either isospin 0, 
($K \bar{K}, \eta \eta'$), or isospin 1, ($\pi \pi, \pi \eta,K \bar{K},\pi \eta'   $).

For each partial-wave $l$ unitarity imposes a restriction on
the (inverse) $T$-matrix above the pertinent thresholds
\begin{equation} \label{unit}
\mbox{Im} T^{-1}_l = - \frac{|\mbox{\boldmath$q$}_{cm}|}{8 \pi \sqrt{s}}
\end{equation}
with $\sqrt{s}$ and
$\mbox{\boldmath$q$}_{cm}$ being the energy and the three-momentum 
of the particles in the center-of-mass frame of the channel under consideration, respectively.
Hence, the imaginary part of $T^{-1}_l$ is equal to the imaginary piece
of the fundamental scalar loop integral $G_{m \bar{m}}$
above threshold, Eq.~(\ref{eq:g}).

Following the work of \cite{OM,Inoue} in the baryonic sector, we will
adjust the real piece of $G_{m \bar{m}}$ by
introducing a scale dependent constant for each channel in analogy
to a subtraction constant of a dispersion relation for $T^{-1}_l$.
This will compensate the regularization scale dependence of $G_{m \bar{m}}$
and bring our results into better agreement with experimental data.
In a more general way, one could model the real parts by taking any analytic
function in $s$ and the baryon and meson masses.
This option has been successfully applied for the case of $SU(2)$ baryon
ChPT in \cite{NA}, but is not necessary in the present work, as we can reproduce
the experimental phase shifts for $s$- and $p$-wave meson-meson scattering
already very accurately with the first method.

\begin{figure}
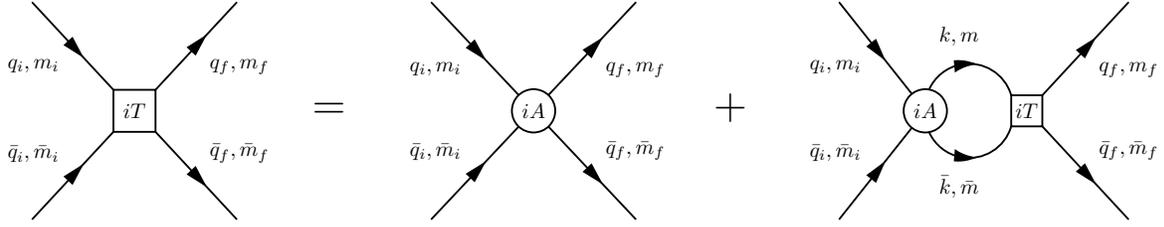

\centering
\begin{tabular}{ccccc}
\parbox{0.19\textwidth}{\includegraphics[scale=0.8]{feynps.5}} & \quad {\Large =} & \quad
\parbox{0.19\textwidth}{\includegraphics[scale=0.8]{feynps.6}} & \quad {\Large +} & \quad
\parbox{0.27\textwidth}{\includegraphics[scale=0.8]{feynps.9}}
\end{tabular}
\caption{Diagrammatic illustration of the  
         Bethe-Salpeter equation for meson-meson rescattering. The center-of-mass
         momentum is denoted by $\sqrt{s} = q_i + \bar{q}_i =  q_f + \bar{q}_f = k + \bar{k}$.}
\label{fig3a}
\end{figure}

The inverse of the amplitude $T^{-1}$ can be decomposed into real
and imaginary parts \footnote{For brevity we suppress the subscript $l$.}
\begin{equation} \label{invers}
T^{-1} = \tau^{-1} + \tilde{G}
\end{equation}
with
\beq   \label{subconst}
\tilde{G}_{m \bar{m}} = G_{m \bar{m}} (\mu) + a_{m \bar{m}} (\mu) ,
\eeq
where $\tau^{-1}$ and Re$[\tilde{G}]$ give the real part and Im$[\tilde{G}]$ gives
the imaginary part of $T^{-1}$ as required by unitarity, Eq.~(\ref{unit}).
The scale dependences of $G$ and $a$ on $\mu$ cancel each other and
we will choose the constant $a$ to depend on the angular momentum $l$.
Inverting  (\ref{invers}) yields
\begin{equation} \label{tau}
T = [\mathds{1} + \tau \cdot \tilde{G}]^{-1} \; \tau
\end{equation}
which is understood to be a matrix equation that couples the different channels. 
The matrix $\tilde{G}$ is diagonal
and includes the expressions for the loop integrals in each channel.
Expanding expression (\ref{tau})
\begin{equation}
T = \tau - \tau \cdot \tilde{G} \cdot \tau \ldots
\end{equation}
and matching the first term in the expansion
to our tree level amplitude for each partial-wave
\begin{equation}
\tau_l = A_l 
\end{equation}
our final expression for the $T$ matrix reads 
\begin{equation} \label{eq:tmat}
T = [\mathds{1} + A \cdot \tilde{G}]^{-1} \; A
\end{equation}
which amounts to a summation of a bubble chain in the $s$-channel.
This is equivalent to a Bethe-Salpeter equation with $A$ as
potential, see Fig.~\ref{fig3a}.

The unknown LECs of the chiral Lagrangian must be fitted
to experimental data. This has partially been accomplished in \cite{BB2},
where after applying the same approach as in the present investigation
we constrained the LECs of the Lagrangian up to fourth chiral order by comparing
the results with the experimental $s$-wave phase shifts of meson-meson scattering.
Agreement was achieved in the isospin $I=0, \sfrac{1}{2}$ channels up to 1.2 GeV and in the isospin
$I=\sfrac{3}{2}, 2$ channels up to 1.5 GeV. 
However, variations of some of the parameters which have been set to zero for simplicity
do not yield any significant effect for the $s$-wave phase shifts. In fact, some of them could
be constrained from the hadronic decays of the $\eta$ and $\eta'$ 
\cite{BB3}.
A good fit to the decays \cite{BB3} and the $s$-wave scattering data \cite{BB2}
is given by
\begin{equation}\label{eq:par2}
\begin{array}{lll}
\cvtwid{2}{1}= \cvtwid{2}{2}= 0 ,&&
\\
\cbeta{0}{0}=0.56\times 10^{-3},&\quad&
\cbeta{3}{0}=-0.3\times 10^{-3},
\\
\cbeta{5}{0}=1.4\times 10^{-3},&\quad&
\cbeta{6}{0}=0.06\times 10^{-3}
\end{array}
\end{equation}
with $\cvtwid{2}{2}= \sfrac{1}{4}\decay^2-\sqrt{6}\coeffv{3}{1}-3\coeffv{2}{2}$
and all the remaining parameters being zero. 
It is not trivial that with such a small number
of parameters we have been able to explain a variety of data within the approach.

\begin{figure}
\centering
\begin{overpic}[width=0.4\textwidth]{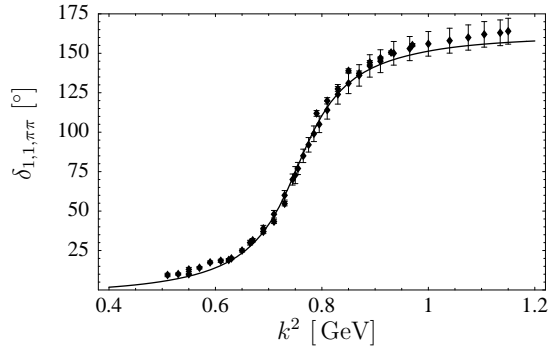}
\put(45,-4){\scalebox{0.8}{$k^2 \ [\GeV]$}}
\put(-7,27){\rotatebox{90}{\scalebox{0.8}{$\delta_{1,1,\pi \pi} \ [^{\circ}]$}}}
\end{overpic}
\caption{Fit to the experimental phase shifts (\cite{Pro, EM}) for $\pi \pi \to \pi \pi$
         in the $I = J = 1$ channel.}
\label{fig4}
\end{figure}

In the present investigation, we are particularly interested in the $p$-wave amplitudes
which will be the only contributions to the photonic decays.
Our fit to the scattering data from \cite{Pro} and \cite{EM}
is shown in Fig.~\ref{fig4}.
We are able to use the same set of parameters as for the $s$-wave scattering in \cite{BB2}
with only one non-vanishing scale dependent constant
\begin{equation}
a_{m_{\pi} m_{\pi}} = - 4.37\times 10^{-2}
\end{equation}
at $\mu = 1.0 \GeV$ and we take $\mu = 1.0 \GeV$ in all channels.
It is also important to note, that the parameter choice in Eq.~(\ref{eq:par2})
is not unique, since
variations in one of the parameters may be compensated by the other ones.
Nevertheless, we prefer to work with this choice; it is capable of describing
the experimental phase shifts both for $s$- and $p$-waves
with a small set of parameters and motivated
by the assumption that most of the OZI-violating and 1/$N_c$-suppressed parameters are not important,
although $\beta_6^{(0)}$ and $\coeffv{3}{1}, \coeffv{2}{2}$ in $\cvtwid{2}{1}, \cvtwid{2}{2}$
have small but non-vanishing values.

\subsection{Meson-meson rescattering in  $\mbox{\boldmath$\pi^0, \eta ,\eta' 
  \rightarrow \gamma \gamma$}$}\label{sec:rescat}
\begin{figure}
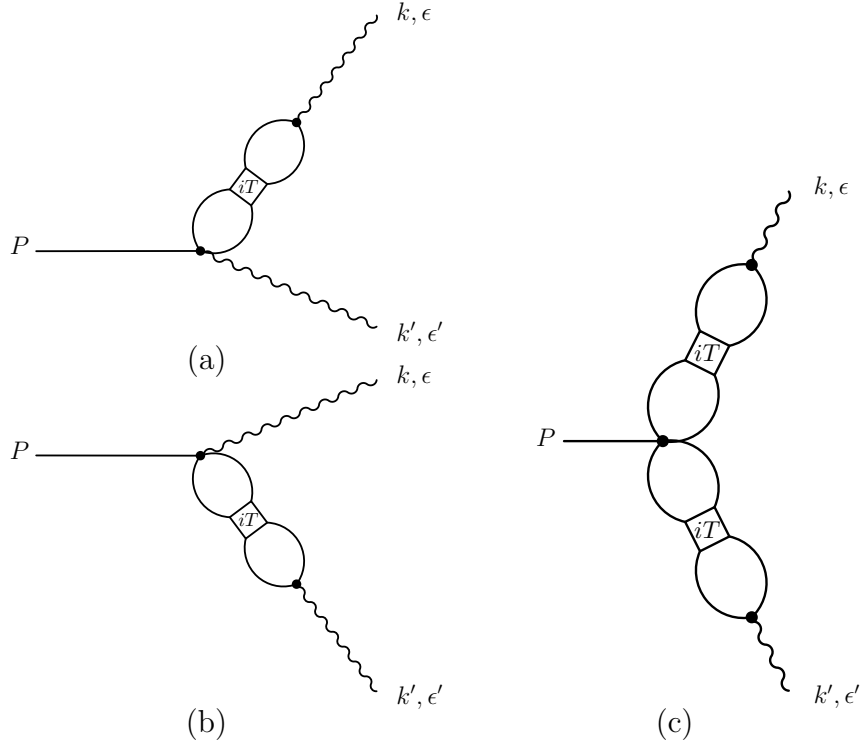

\centering
\begin{minipage}[b]{0.3\textwidth}
\centering
\begin{overpic}[width=0.9\textwidth]{feynps.10}
\put(-7,22){\scalebox{0.8}{$P$}}
\put(105,90){\scalebox{0.8}{$k, \epsilon$}}
\put(105,-3){\scalebox{0.8}{$k', \epsilon'$}}
\end{overpic} \\
(a) \\
\begin{overpic}[width=0.9\textwidth]{feynps.11}
\put(-7,68){\scalebox{0.8}{$P$}}
\put(105,90){\scalebox{0.8}{$k, \epsilon$}}
\put(105,-3){\scalebox{0.8}{$k', \epsilon'$}}
\end{overpic} \\
(b)
\end{minipage}
\hspace{0.1\textwidth}
\begin{minipage}[b]{0.2\textwidth}
\centering
\begin{overpic}[width=0.9\textwidth]{feynps.12}
\put(-5,49){\scalebox{0.8}{$P$}}
\put(50,99){\scalebox{0.8}{$k, \epsilon$}}
\put(50,-2){\scalebox{0.8}{$k', \epsilon'$}}
\end{overpic} \\
(c)
\end{minipage}
\caption{Set of meson-meson rescattering processes in the decay $P \to \gamma^{(*)} \gamma^{(*)}$
         included in our approach.}
\label{fig5}
\end{figure}

The meson-meson rescattering processes from the previous section can be employed as
effective interaction kernels for two mesons in the decay process, see Fig.~\ref{fig5}.
In order to perform the remaining loop integrations in Fig.~\ref{fig5}
which are not included in the effective two-meson interaction kernel $iT$, 
we rewrite the partial-wave decomposition of the $T$ matrix 
of the meson-meson scattering process as follows \cite{BB2}.
The general expression for $T$ depends only on scalar combinations
of the momenta which can be expressed in terms of the Mandelstam variables.
The Mandelstam invariants $s$, $t$ and $u$ are defined as the center-of-mass
energy squared $s=(q_i+\bar q_i)^2=p^2$, the momentum transfer squared
$t=(q_i-q_f)^2$ and the crossed momentum transfer squared
$u=(q_i-\bar q_f)^2$, where we used the notation of the previous section, see Fig.~\ref{fig3a}.
The constraint $s+t+u=q_i^2+\bar q_i^2+q_f^2+\bar q_f^2=m_i^2+\bar m_i^2+m_f^2+\bar m_f^2$
allows one to neglect the combination $t+u$ in favor of $t-u$ and
the scalar amplitude can be written as $T(s,t-u)$.
Since we restrict ourselves to the effective Lagrangian up to fourth chiral order,
the partial-wave decomposition of $T$---which is the expansion of $T$ in $t-u$---is given by 
\begin{equation}\label{eq:ResPartialWave}
T=\tsum\nolimits_l \hat{T}_l J_l=\hat{T}_s J_s+\hat{T}_p J_p+\hat{T}_d J_d ,
\end{equation}
where the partial-wave operator $J_l$
is a polynomial of
degree $l$ in $t-u$ and proportional
to the Legendre polynomial $P_l$ in the cosine of the scattering angle.
The $J_l$ can be written as
{\arraycolsep0pt\begin{eqnarray}
J_s=&&\mathrel{}1,\nonumber\\
J_p=&&\mathrel{}h_{\mu\nu}q_i^\mu q_f^\nu
=\frac{t-u}{4}+\frac{(q_i^2-\bar q_i^2)(q_f^2-\bar q_f^2)}{4s},
\nonumber\\
J_d=&&
J_p^2-\frac{h_{\mu\nu}q_i^\mu q_i^\nu\,h_{\rho\sigma}q_f^\rho q_f^\sigma}{3} ,
\end{eqnarray}}%
with
\begin{equation}
h_{\mu\nu}=-g_{\mu\nu}+p_\mu p_\nu/p^2.
\end{equation}
It is now straightforward to perform the remaining loop calculations
in Fig.~\ref{fig5}, since the $\hat{T}_l$ depend only on the Mandelstam variable
$s$, {\it i.e.} the momentum squared of one of the photons. Hence, they are
independent of the loop momenta and can be factored out of the loop integral.
The remaining pieces $J_l$ can be expressed in terms of the loop momenta
and the loop integrations can be performed.
In this way, the $T$ matrix is treated as an effective vertex, which summarizes
the infinite chain of meson-meson rescattering processes.
One checks explicitly that $s$- and $d$-waves drop out,
whereas only the $t-u$ piece in $J_p$ contributes.
Only the channels ($\pi \pi, K \bar{K}$)
contribute so that $q_f^2 = \bar{q}_f^2$, and hence the limit $s \rightarrow 0$
for on-shell photons does not yield a divergence in $J_p$.
The results for the diagrams of Fig.~\ref{fig5} read
\begin{equation} \label{oneexc}
\begin{array}{l}
\mathcal{A}^{\textit{(1\,CC)}}(P \to \gamma^{(*)} \gamma^{(*)}) = 
e^2 k_\mu \epsilon_\nu k'_\alpha \epsilon'_\beta \epsilon^{\mu \nu \alpha \beta } 
\frac{1}{4 \pi^2 F_{P}^3} 
\sum_{a}' \gamma_{P}^{\textit{(1\,CC)},a} \\[2ex]
\qq \times \left[ \tilde{I}(m_{a}^2;k^2) [\hat{T}_{p}^{(a \to \pi^\pm)}(k^2) \tilde{I}(m_{\pi}^2;k^2)
+ \hat{T}_{p}^{(a \to K^\pm)}(k^2) \tilde{I}(m_{K}^2;k^2)] \right. \\[2ex]
\qq \quad \left. + \tilde{I}(m_{a}^2;k'^2)  [\hat{T}_{p}^{(a \to \pi^\pm)}(k'^2) 
\tilde{I}(m_{\pi}^2;k'^2) + \hat{T}_{p}^{(a \to K^\pm)}(k'^2) \tilde{I}(m_{K}^2;k'^2)] \right]
\end{array}
\end{equation}
for the one vector-meson exchange and
\begin{equation} \label{twoexc}
\begin{array}{l}
\mathcal{A}^{\textit{(2\,CC)}}(P \to \gamma^{(*)} \gamma^{(*)}) =
e^2 k_\mu \epsilon_\nu k'_\alpha \epsilon'_\beta \epsilon^{\mu \nu \alpha \beta }
\frac{1}{4 \pi^2 F_{P}^5}
\sum_{a,b}' \gamma_{P}^{\textit{(2\,CC)},a,b} \\[2ex]
\qq \times \ \tilde{I}(m_{a}^2;k^2) [\hat{T}_{p}^{(a \to \pi^\pm)}(k^2) \tilde{I}(m_{\pi}^2;k^2)
+ \hat{T}_{p}^{(a \to K^\pm)}(k^2) \tilde{I}(m_{K}^2;k^2)] \\[2ex]
\qq \times \ \tilde{I}(m_{b}^2;k'^2) [\hat{T}_{p}^{(b \to  \pi^\pm)}(k'^2) \tilde{I}(m_{\pi}^2;k'^2)
+ \hat{T}_{p}^{(b \to K^\pm)}(k'^2) \tilde{I}(m_{K}^2;k'^2)]
\end{array}
\end{equation}
for the two vector-meson exchange, where $\hat{T}_{p}^{(a \to b)}$ denotes the $p$-wave amplitude
in Eq.~(\ref{eq:ResPartialWave}) for channel $a$ scattering into channel $b$. The symbol $\sum'$ 
indicates summation over the meson pairs $\pi^+ \pi^-$, $K^+ K^-$ and $K^0 \bar{K}^0$.
Note that the exchange of two vector mesons
arises due to the five-meson vertex in the piece $\langle \Sigma^5 \rangle$ of the WZW action,
Eq.~(\ref{eq:wzw}).
The coefficients in Eq.~(\ref{oneexc}) are given by
\begin{equation}
\begin{array}{l}
\gamma_{\pi^0}^{\textit{(1\,CC)},\pi^{\pm}} = \gamma_{\pi^0}^{\textit{(1\,CC)},K^{\pm}}
= \gamma_{\pi^0}^{\textit{(1\,CC)},K^0 \bar{K}^0} = - \dfrac{1}{2} , \\[2ex]
\gamma_{\eta}^{\textit{(1\,CC)},\pi^{\pm}} = \gamma_{\eta}^{\textit{(1\,CC)},K^{\pm}}
= - \dfrac{1}{6} \left[\sqrt{3} + \dfrac{4\sqrt{2}}{3}(m_{K}^2 - m_{\pi}^2) 
\dfrac{\cvtwid{2}{1}}{\coeffv{0}{2}} (\sqrt{6} - 48 \pi^2 w_{3}^{(1)}) \right] , \\[2ex]
\gamma_{\eta}^{\textit{(1\,CC)},K^0 \bar{K}^0} = \dfrac{\sqrt{3}}{2} , \qq
\gamma_{\eta'}^{\textit{(1\,CC)},K^0 \bar{K}^0} = 2 \sqrt{\dfrac{2}{3}}
(m_{K}^2 - m_{\pi}^2) \left(- \dfrac{\cvtwid{2}{1}}{\coeffv{0}{2}}
+ 4 \beta_{5,18} \right) , \\[2ex]
\gamma_{\eta'}^{\textit{(1\,CC)},\pi^{\pm}} = \gamma_{\eta'}^{\textit{(1\,CC)},K^{\pm}}
= - \dfrac{1}{6} \left[\sqrt{6} - 48 \pi^2 w_{3}^{(1)} + \dfrac{4\sqrt{6}}{3}
(m_{K}^2 - m_{\pi}^2) \left(- \dfrac{\cvtwid{2}{1}}{\coeffv{0}{2}} 
+ 4 \beta_{5,18} \right) \right] . \\
\end{array}
\end{equation}
The coefficients $\gamma_{P}^{\textit{(2\,CC)},a,b}$ for the case of two coupled channels 
are symmetric under $a \leftrightarrow b$. The non-vanishing ones read
\begin{equation}
\begin{array}{lclclcl}
\gamma_{\pi^0}^{\textit{(2\,CC)},\pi^{\pm},K^{\pm}} 
& = & \gamma_{\pi^0}^{\textit{(2\,CC)},\pi^{\pm},K^0 \bar{K}^0} & = & \dfrac{3}{4} , \\[2ex]
\gamma_{\eta}^{\textit{(2\,CC)},\pi^{\pm},K^{\pm}}
& = & - \gamma_{\eta}^{\textit{(2\,CC)},\pi^{\pm},K^0 \bar{K}^0}
& = & - \dfrac{1}{2} \gamma_{\eta}^{\textit{(2\,CC)},K^{\pm},K^0 \bar{K}^0} 
& = & \dfrac{\sqrt{3}}{4} , \\[2ex]
\gamma_{\eta'}^{\textit{(2\,CC)},\pi^{\pm},K^{\pm}}
& = & - \gamma_{\eta'}^{\textit{(2\,CC)},\pi^{\pm},K^0 \bar{K}^0}
& = & - \dfrac{1}{2} \gamma_{\eta'}^{\textit{(2\,CC)},K^{\pm},K^0 \bar{K}^0}
& = & \sqrt{\dfrac{2}{3}} (m_{K}^2 - m_{\pi}^2) \left(- \dfrac{\cvtwid{2}{1}}{\coeffv{0}{2}}
+ 4 \beta_{5,18} \right) .
\end{array}
\end{equation}
The loop integral $\tilde{I}$ in Eqs.~(\ref{oneexc},\ref{twoexc}) is given by
\begin{equation} \label{eq:int2}
\tilde{I}(m^2;p^2) = I(m^2;p^2) + C_P \ p^2 
\end{equation}
where the integral $I$ is defined in Eq.~(\ref{eq:int1}).
Here we made use of the freedom to take arbitrary values for the analytic pieces
of the integrals which corresponds to a specific choice
of counterterm contributions. To be more precise, we have kept in the pion loops
a term of the type $p^2$ ($C_\pi = - \frac{1}{12 \pi^2}$), 
while neglecting all other analytic portions. 
Alternatively, one could have altered the regularization 
scale $\mu$ of the integrals relevant for the $p^2$ term, 
but we preferred to take explicit analytic pieces, while keeping $\mu = 1\GeV$
fixed in all channels, which is 
similar to adding a subtraction constant as in Eq.~(\ref{subconst}).
As we will see in the next section, the inclusion of such analytic portions which are beyond
the accuracy of the one-loop calculation yields an improved fit to the experimental data
and accounts for dynamical effects of higher chiral order.

Clearly, we are missing further unitarity corrections beyond the one-loop calculation.
However, from the discussion below it will become clear that the set of diagrams
included in our model is consistent with available data.

\section{Numerical results}\label{sec:results}
In this section, we will discuss the numerical results of our calculation.
In order to compare our results with experimental data, we include
the counterterm contributions from Eq.~(\ref{ctcontr}).
At the one-loop level, it was not necessary to take them into account, as
the one-loop formulae could already be brought to agreement with the 
experimental decay widths by fitting either $F_{\eta'}$ and $\cvtwid{2}{1}$ or
$\cvtwid{2}{1}$ and one of the contact terms $w_1^{(1)}$, $w_2^{(1)}$, $w_3^{(1)}$.
The inclusion of the coupled channel analysis, on the other hand, leads to changes in the decay amplitude
for two on-shell photons which must be compensated by counterterms
that are proportional to the quark mass matrix, {\it i.e.} those with the coefficients
$\bar{w}_{3}^{(0)}$, $\bar{w}_{4}^{(0)}$, $\bar{w}_{5}^{(1)}$ and $\bar{w}_{6}^{(1)}$.
It turns out that the inclusion of counterterms of sixth chiral order as discussed
in Section \ref{sec:wave} is sufficient to compensate these additional contributions.
Furthermore, $k^2$ dependent counterterms 
($\bar{w}_{1}^{(0)}$, $\bar{w}_{2}^{(1)}$)
are needed to obtain agreement
with data for space-like photons with squared four-momenta $k^2 <0$, since the non-analytic contributions
from the one-loop diagrams are too small to account for the behavior of the
transition form factor in the space-like region, where furthermore 
contributions from the BSE are almost negligible.
The two possible terms from the sixth order Lagrangian $\mathcal{L}_{ct}^{(6)}$, Eq.~(\ref{lagrct}),
are sufficient to bring our results into better agreement with the data
for photon virtualities up to $k^2 \approx -0.8$ GeV$^2$.
The values of the counterterms are (in units of GeV$^{-2}$)
\begin{equation}
\begin{array}{lcllcllcl}
\bar{w}_{1}^{(0)} & = & -5.5 \times 10^{-3} , \qq &
\bar{w}_{2}^{(1)} & = & -1.5 \times 10^{-3} , \qq &
\bar{w}_{3}^{(0)} & = &  1.31 \times 10^{-2} , \\
\bar{w}_{4}^{(0)} & = & -1.4 \times 10^{-3} , \qq &
\bar{w}_{5}^{(1)} & = & -0.47 \times 10^{-3} , \qq &
\bar{w}_{6}^{(1)} & = & -0.16 \times 10^{-3} .
\end{array}
\end{equation}
The parameters $\bar{w}_{5}^{(1)}$ and $\bar{w}_{6}^{(1)}$ enter only in the combination
$\frac{4}{3}(m_{K}^2 + 2 m_{\pi}^2) \bar{w}_{5}^{(1)}
+ 4(2 m_{K}^2 + m_{\pi}^2) \bar{w}_{6}^{(1)}$, see Eq.~(\ref{eq:ctcoeff}), hence their values cannot
be fixed separately. Since the $\bar{w}_{6}^{(1)}$ term is $1/N_c$ suppressed, we choose 
$\bar{w}_{6}^{(1)} = \bar{w}_{5}^{(1)} / 3$.

\begin{figure}
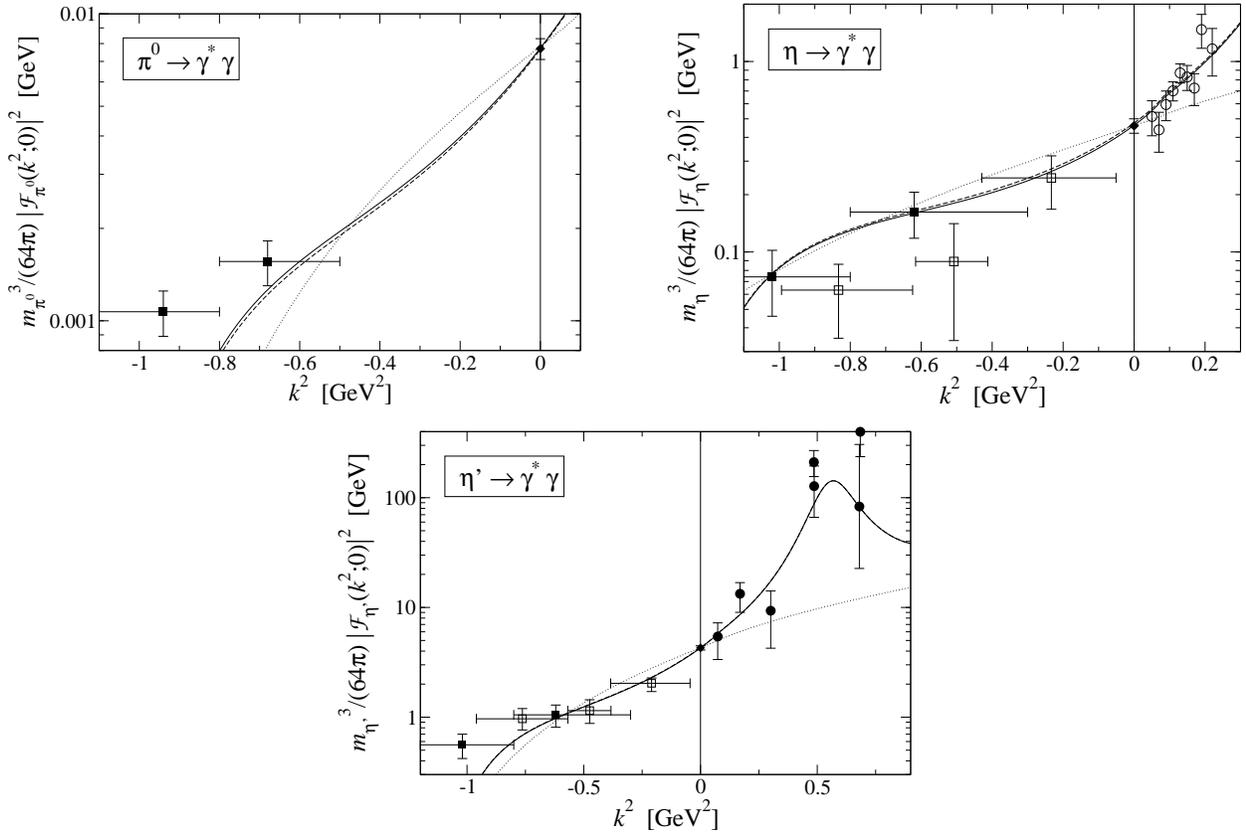

\centering
\begin{minipage}{0.45\textwidth}
\includegraphics[width=1.0\textwidth]{fitpi0.eps}
\end{minipage}
\hspace{0.05\textwidth}
\begin{minipage}{0.45\textwidth}
\includegraphics[width=1.0\textwidth]{fiteta.eps}
\end{minipage} \\[1ex]
\begin{minipage}{0.45\textwidth}
\includegraphics[width=1.0\textwidth,clip]{fitetap.eps}
\end{minipage}
\caption{Results of the one-loop calculation (dotted), including the diagrams Figs.~\ref{fig5}a+b (dashed),
         full calculation (solid).
         Data: solid squares \cite{CELLO}, open squares \cite{TPC}, diamond \cite{pdg}, 
         open circles \cite{LG2}, solid circles \cite{LG1}.}
\label{fig6}
\end{figure}

The results of our model are compared in Fig.~\ref{fig6} 
with the transition form factors for the decay of the $\pi^0$, $\eta$ 
and $\eta'$ into one on-shell and one off-shell photon.
The transition form factor $\mathcal{F}_P(k^2;k'^2)$ is defined as
\begin{equation}
\mathcal{A}(P \to \gamma^{*} \gamma^{(*)}) = k_{\mu} \epsilon_{\nu} k'_{\alpha} \epsilon'_{\beta} 
\epsilon^{\mu \nu \alpha \beta} \mathcal{F}_P(k^2;k'^2)
\end{equation}
and we plot the quantity $\frac{m_{P}^3}{64 \pi} |\mathcal{F}_P(k^2;k'^2=0)|^2$ which yields 
the width of the decay into two physical photons at $k^2 = 0$.

The overall good agreement with the data indicates that our model is capturing
the important physics.
In our fit, we have set the non-anomalous terms of unnatural parity at fourth chiral order
to zero, $w_1^{(1)} = w_2^{(1)} = w_3^{(1)} = 0$, since a good fit to the data
can already be achieved without them, while keeping $F_{\eta'}$ fixed at $F_{\eta'} = 1.1 \; F_\pi$.
Furthermore, we set the mixing parameter $\cvtwid{2}{1}$ to zero
which is consistent with a previous coupled channel calculation on hadronic decays of the
$\eta$ and $\eta'$ \cite{BB3}. 
It should be emphasized that for a non-perturbative coupled channel approach
the values of the coupling constants do not necessarily coincide with those
from a perturbative loop-expansion, which was already pointed out in \cite{BB3}.
It is therefore not surprising that $\cvtwid{2}{1}$ and the $w_i^{(1)}$ differ
in both schemes.
The inclusion of the diagrams in Figs.~\ref{fig5}a,~\ref{fig5}b which contain only one 
coupled channel and thus are associated with the exchange of only one vector-meson already yields the 
crucial structure of the curves (dashed line). The exchange of two vector-mesons (Fig.~\ref{fig5}c) 
is merely a small correction which could even be compensated by a change of the parameters of the 
$\mathcal{O}(p^6)$ counterterms.
For completeness we have also shown the numerical results from the one-loop calculation
which have been supplemented by the $k^2$ dependent counterterms at sixth chiral order,
in order to be in better agreement for $k^2 <0$. Due to this procedure the $k^2$ dependent terms
differ from those in the coupled channel approach.
Nevertheless, with this choice of parameters it is not possible to reproduce the sharp
increase of the form factor for time-like photons, $k^2 > 0$, at the one-loop level, and we miss 
in any case the resonance structure in the $\eta'$ decay.
 
\begin{figure}
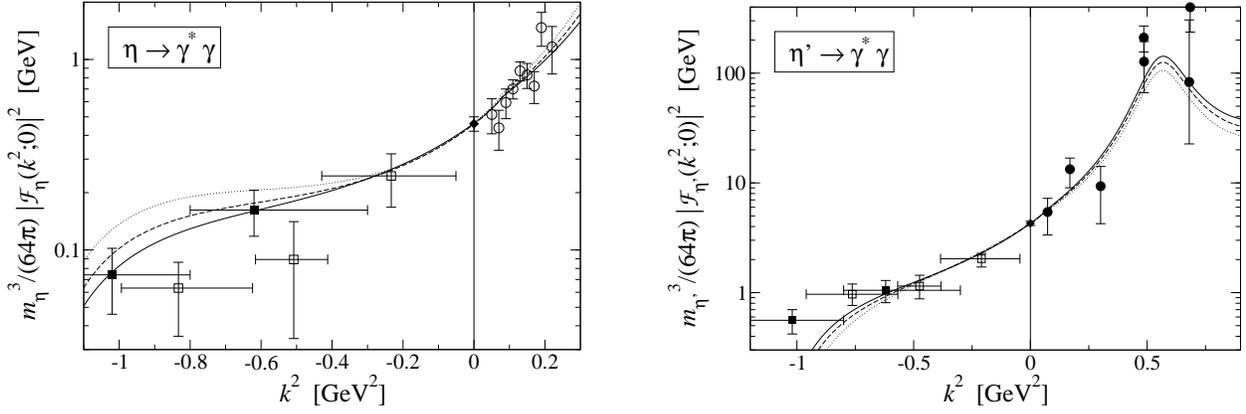

\centering
\begin{minipage}{0.45\textwidth}
\includegraphics[width=1.0\textwidth,clip]{etav21.eps}
\end{minipage}
\hspace{0.05\textwidth}
\begin{minipage}{0.45\textwidth}
\includegraphics[width=1.0\textwidth,clip]{etapv21.eps}
\end{minipage}
\caption{Dependence of $\mathcal{F}_{\eta}$ and $\mathcal{F}_{\eta'}$ 
         on the mixing parameter $\cvtwid{2}{1}$: 
	 $\cvtwid{2}{1} = 0$ (solid), $\cvtwid{2}{1} = 0.5 F_{\pi}^{2}/4$ (dashed), 
	 $\cvtwid{2}{1} = 1.15 F_{\pi}^{2}/4$ (dotted).}
\label{fig6a}
\end{figure}

The dependence of our results on the $\eta$-$\eta'$ mixing parameter $\cvtwid{2}{1}$ is depicted 
in Fig.~\ref{fig6a}. We have plotted the curves for different values 
of $\cvtwid{2}{1}$, which indicate that a value for $\cvtwid{2}{1}$ around zero
is in better agreement with the data. 
The results for $\pi^0 \to \gamma^* \gamma$ are, of course, not affected by changing $\cvtwid{2}{1}$.

\begin{figure}
\centering
\includegraphics[width=0.45\textwidth,clip]{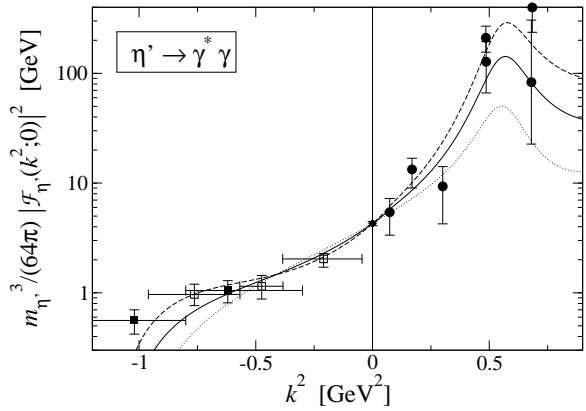}
\caption{Dependence of $\mathcal{F}_{\eta'}$ on the LEC $w_{3}^{(1)}$: 
         $w_{3}^{(1)} = -3.0 \times 10^{-3}$ (dashed), $w_{3}^{(1)} = 0$ (solid), 
	 $w_{3}^{(1)} = 3.0 \times 10^{-3}$ (dotted).}
\label{fig6b}
\end{figure}

In Fig.~\ref{fig6b} we show our results for different values of the LEC $w_{3}^{(1)}$. When keeping 
$\cvtwid{2}{1} = 0$ only the process $\eta' \to \gamma^* \gamma$ is altered. From the plot it can be seen 
that $w_{3}^{(1)}$ has significant influence on the peak structure at $k^2 \approx 0.6 \GeV^2$ and 
we can exclude values $w_{3}^{(1)} > 2.0 \times 10^{-3}$.

Another quantity of interest is the slope parameter of the transition
form factor. It is defined as the logarithmic derivative of the form factor $\mathcal{F}_P$
at $k^2 = 0$
\begin{equation}
b_P = \left. \frac{d}{d k^2} \ln{\mathcal{F}_P(k^2;k'^2=0)} \right|_{k^2=0} .
\end{equation}
In Table~\ref{table1} we compare our values with simple pole fits
to the form factors. The value for $\pi^0$ 
is the world average given in \cite{pdg}. For $\eta$ and $\eta'$ we cite three measurements, 
the Lepton-G Collaboration \cite{LG2, LG1}, the TPC/2$\gamma$ Collaboration \cite{TPC} and the CELLO 
Collaboration \cite{CELLO}. From a pole fit a characteristic mass $\Lambda_P$ is extracted 
and related to the slope parameter via $b_P = 1/\Lambda_{P}^2$. The obtained values are given by
$\Lambda_{\eta} = (0.72 \pm 0.09) \GeV$ \cite{LG2}, 
$\Lambda_{\eta} = (0.70 \pm 0.08) \GeV$ \cite{TPC}, 
$\Lambda_{\eta} = (0.84 \pm 0.06) \GeV$ \cite{CELLO} and 
$\Lambda_{\eta'} = (0.77 \pm 0.18) \GeV$ \cite{LG1}, 
$\Lambda_{\eta'} = (0.85 \pm 0.07) \GeV$ \cite{TPC}, 
$\Lambda_{\eta'} = (0.79 \pm 0.04) \GeV$ \cite{CELLO}.
With one exception we achieve good agreement within error bars with the pole fits
to the form factors. We point out that in the 
determination of the given errors for $b_P$ it has not been taken into account that in principle a different 
ansatz for the form factor can lead to different values of the slope parameter.
In our model a significant part of the slope 
parameters is induced by the 
meson-meson rescattering processes. 
Again the exchange of two vector-mesons plays  a negligible role.

\begin{table}
\centering
\begin{tabular}{c|cc|c|c|c}
& Experiment & & NLO & 1 CC & full \\
\hline
$\pi^0$ & $1.76 \pm 0.22$ & \cite{pdg} & 1.19 & 1.98 & 1.95 \\
\hline
        & $1.9  \pm 0.4$  & \cite{LG2} & & & \\
$\eta$  & $2.0  \pm 0.5$  & \cite{TPC} & 0.73 & 1.57 & 1.58 \\ 
	& $1.42 \pm 0.21$ & \cite{CELLO} & & & \\
\hline
        & $1.7  \pm 0.8$  & \cite{LG1} & & & \\
$\eta'$ & $1.38 \pm 0.23$ & \cite{TPC} & 0.85 & 1.78 & 1.79 \\
	& $1.59 \pm 0.18$ & \cite{CELLO} & & & \\
\end{tabular}
\caption{Slope parameters $b_P$ in units of $\GeV^{-2}$ derived from pole fits to the experimental data, 
         the next-to-leading order calculation,
         including one coupled channel (diagrams Fig.~\ref{fig5}a+b), and the full calculation.}
\label{table1}
\end{table}

Within our model, contributions from composed states of three pseudoscalar
mesons which would correspond to the $\omega(782)$ have been neglected.
However, the effects of the $\omega(782)$ should be similar to those of the $\rho(770)$
due to their almost equal masses.
It may well be, that by keeping some of the analytic portions of the loop integrals
in Eqs.~(\ref{oneexc},\ref{twoexc})
which amounts to a particular choice of higher order counterterms, 
we took into account the dynamical effects of the
$\omega(782)$. Within our model, the contributions from the vector-mesons
should thus be regarded to be a combination of both the $\omega(782)$ and the $\rho(770)$.

\begin{figure}
\centering
\begin{minipage}{0.45\textwidth}
\centering
\includegraphics[width=1.0\textwidth,clip]{diagpi0.eps}
\end{minipage}
\hspace{0.05\textwidth}
\begin{minipage}{0.45\textwidth}
\centering
\includegraphics[width=1.0\textwidth,clip]{diageta.eps}
\end{minipage} \\
\begin{minipage}{0.45\textwidth}
\centering
\includegraphics[width=1.0\textwidth,clip]{diagetap.eps}
\end{minipage}
\hspace{0.05\textwidth}
\begin{minipage}{0.45\textwidth}
\centering
\begin{overpic}[width=0.8\textwidth]{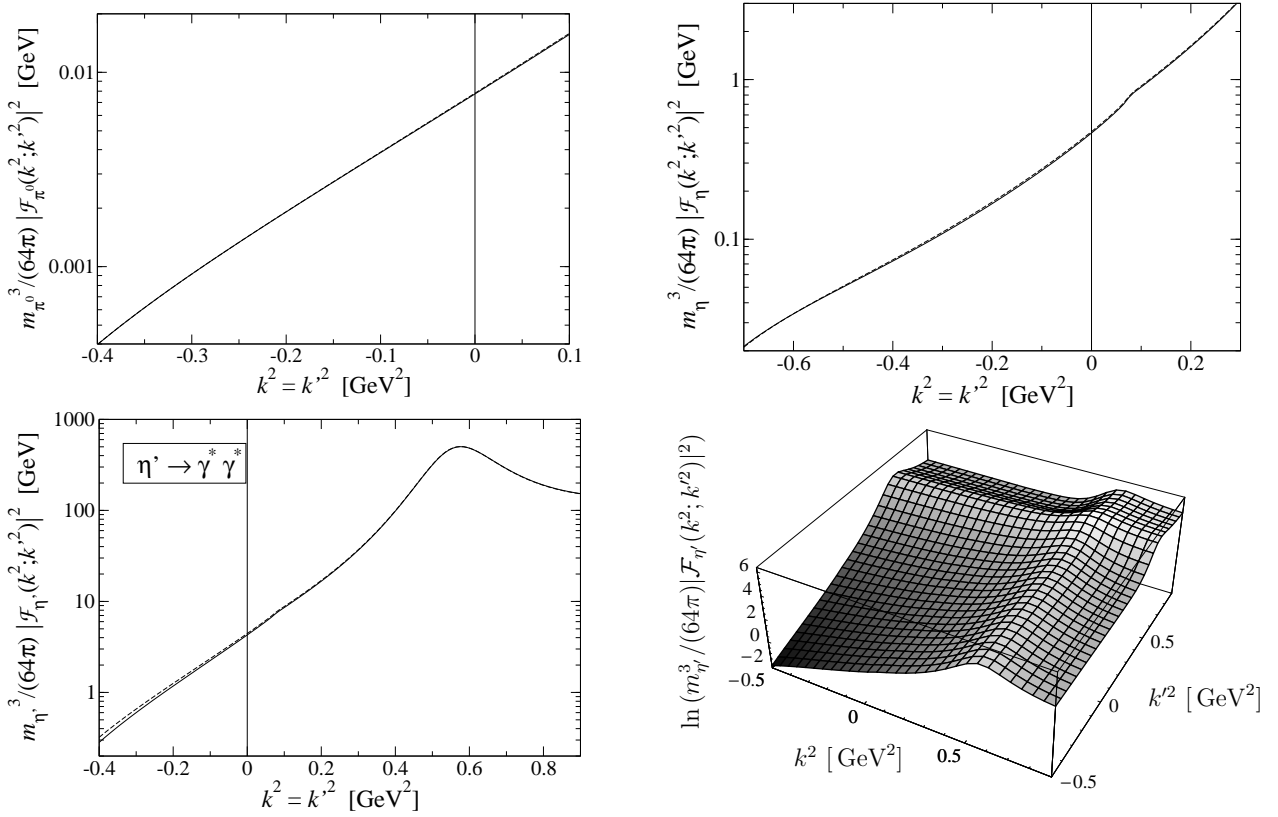}
\put(15,7){\scalebox{0.8}{$k^2 \ [\GeV^2]$}}
\put(92,20){\scalebox{0.8}{$k'^2 \ [\GeV^2]$}}
\put(-10,15){\rotatebox{90}{\scalebox{0.8}{%
             $\ln{(m_{\eta'}^3/(64 \pi) |\mathcal{F}_{\eta'}(k^2;k'^2)|^2)}$}}}
\end{overpic}
\end{minipage}
\caption{Prediction of the transition form factors for two off-shell photons with $k^2 = k'^2$ 
         resulting from the full calculation (solid) and from a calculation without the contribution 
	 from Fig.~\ref{fig5}c with two coupled channels (dashed, almost on top of solid line).
         For $\eta' \to \gamma^* \gamma^*$ we also show the transition form factor from the full 
	 calculation in the $(k^2, k'^2)$-plane.}
\label{fig7}
\end{figure}

The contributions of the two vector-meson exchange diagrams, on the other hand,
are almost negligible, as can be seen from Fig.~\ref{fig6}.
This is in sharp contradistinction to the complete Vector-Meson Dominance picture
where the photons can only couple via the exchange of vector-mesons.
Of course, when coupling to an on-shell photon the pertinent vector-meson propagator
reduces to a vertex, but with both photons off-shell, the two approaches should yield
different predictions.
In Fig.~\ref{fig7} we show our predictions for two off-shell photons for 
$\pi^0, \eta, \eta'$ decays with $k^2=k'^2$ as well as the  
$\eta'$ transition form factor in the $(k^2, k'^2)$-plane.
As in the case with one off-shell photon the result of a calculation with one coupled channel  
is almost identical to the full calculation.
A measurement of the transition form factors for two off-shell photons
could serve as a check of our model and help to clarify, whether the exchange
of only one or rather two vector-mesons occurs in these decays.
For one and two off-shell photons with larger space-like momenta the pion transition 
form factor has been calculated within an instanton model of QCD \cite{Dor}. For small 
negative photon virtualities the results are quite similar to ours.

\section{Conclusions}\label{sec:concl}

In the present work, we have investigated the two-photon decays of
$\pi^0, \eta$ and $\eta'$ within a chiral unitary framework.
To this end, we have supplemented the one-loop calculation
of chiral perturbation theory by a Bethe-Salpeter approach which satisfies
unitarity constraints and generates vector-mesons from composed states of two pseudoscalar mesons.

While the one-loop result is sufficient to achieve agreement with the decay widths
$P \rightarrow \gamma \gamma$ $(P=\pi^0, \eta, \eta')$, the vector-meson exchange 
is crucial to reproduce the sharp increase of the transition form factor for
time-like photons in the decays $P \rightarrow \gamma \gamma^*$.
Our method reproduces also the resonance structure in the transition form
factor of the decay $\eta' \rightarrow \gamma \gamma^*$ at photon virtualities around
$k^2 \approx 0.6$ GeV$^2$.

Furthermore, our approach distinguishes between single and double vector-meson exchange,
the latter one being the only contribution in the complete Vector-Meson Dominance picture.
Our study suggests that for the decays with off-shell photons the exchange of
one vector-meson is the dominant contribution, whereas the two vector-meson exchange
is almost negligible which is in contradistinction to complete Vector-Meson Dominance.
The available data on electromagnetic transition form factors is restricted
to decays with exactly one off-shell photon.
When coupling to an on-shell photon, the pertinent vector-meson propagator reduces
to a vertex so that no clear distinction can be made between the two approaches.
However, an experiment with two off-shell photons should be able to clarify,
whether one or two vector-meson exchange occurs.
The question whether double Vector-Meson Dominance holds is also an important issue
for kaon decays and the anomalous magnetic moment of the muon.
We have presented predictions for the decays with two off-shell photons, and
again the exchange of two vector-mesons plays a minor role.

Within our model, contributions from composed states of three pseudoscalar
mesons which would correspond to the $\omega(782)$ have been neglected.
However, the effects of the $\omega(782)$ should be similar to those of the $\rho(770)$.
It may well be, that by keeping some of the analytic portions of the loop integrals
in our coupled channel analysis
which amounts to a particular choice of higher order counterterms, 
we took into account the dynamical effects of the
$\omega(782)$. Within our model, the contributions from the vector-mesons
should thus be regarded to be a combination of both the $\omega(782)$ and the $\rho(770)$.

The present approach can be applied in a straightforward manner to the anomalous
decays $\eta, \eta' \rightarrow \pi^+ \pi^- \gamma$ and the radiative
decays $\eta ,\eta'  \rightarrow \pi^0 \gamma \gamma$
which will be investigated in future studies.

\section*{Acknowledgements}

We are grateful to N. Beisert and B. M. K. Nefkens for useful discussions.

\appendix

\section{Counterterms of order $\mathcal{O}(p^6)$}\label{app:a}

In this section, we discuss the relevant counterterms of order $\mathcal{O}(p^6)$. For the 
covariant derivative and the field strength tensors we use the following notation
\begin{equation} 
\begin{array}{rcl}
\cder_{\mu} U & = & \partial_{\mu} U - i \tilde{r}_{\mu} U + i U \tilde{l}_{\mu} , \\[1ex] 
\tilde{R}_{\mu \nu} & = & \partial_{\mu}\tilde{r}_{\nu} - \partial_{\nu}\tilde{r}_{\mu} 
- i [\tilde{r}_{\mu}, \tilde{r}_{\nu}] , \qq
\tilde{L}_{\mu \nu} = \partial_{\mu}\tilde{l}_{\nu} - \partial_{\nu}\tilde{l}_{\mu} 
- i [\tilde{l}_{\mu}, \tilde{l}_{\nu}] ,
\end{array}
\end{equation}
where $\tilde{r}_{\mu} = v_{\mu} + \tilde{a}_{\mu}$, $\tilde{l}_{\mu} = v_{\mu} - \tilde{a}_{\mu}$ and 
$\tilde a_\mu = a_\mu + \frac{\sqrt{6\lambda}-f}{3f} \langle a_\mu \rangle$. 
The replacement of the original field strength tensors $R_{\mu \nu}$, $L_{\mu \nu}$ by the new 
quantities $\tilde{R}_{\mu \nu}$, $\tilde{L}_{\mu \nu}$ leads to additional counterterms which involve 
the derivative of the singlet component of the axial-vector field, $\langle \partial_\mu a_\nu \rangle$, 
however, such counterterms do not contribute to the processes discussed in the present work and 
are neglected.

In the $SU(3)$ framework there are numerous terms containing six derivatives which according to
\cite{EFS} can be reduced to only one contribution by utilizing methods such as partial integration, 
equation of motion, epsilon relations and Bianchi identities. In our notation it is given by
\begin{equation} \label{eq:app}
\begin{array}{ll}
i\, \bar{W}_1 \epsilon^{\mu \nu \alpha \beta} & \langle (2 U^{\dagger} D^{\lambda} \tilde{R}_{\lambda \mu} U 
+ 2 D^{\lambda} \tilde{L}_{\lambda \mu} + U^{\dagger} \tilde{R}_{\lambda \mu} D^{\lambda} U 
+ D^{\lambda} U^{\dagger} \tilde{R}_{\lambda \mu} U + U^{\dagger} D^{\lambda} U  \tilde{L}_{\lambda \mu} U^{\dagger} 
+ \tilde{L}_{\lambda \mu} D^{\lambda} U^{\dagger} U) \\ 
 & \quad \times \tilde{H}_{\nu \alpha} \, C_{\beta} \rangle ,
\end{array}
\end{equation}
where we made use of the abbreviations
\begin{equation}
\begin{array}{rclrcl}
\tilde{H}_{\mu \nu} & = & U^{\dagger} \tilde{R}_{\mu \nu} U + \tilde{L}_{\mu \nu} , \qq & 
C_{\mu} & = & U^\dagger \cder_\mu U, \\
M & = & U^\dagger \chi+\chi^\dagger U, \qq & 
N & = & U^\dagger \chi-\chi^\dagger U.
\end{array}
\end{equation}
For the decays into two photons Eq.~(\ref{eq:app}) yields the contribution (in the differential form 
notation of \cite{KL1})
\begin{equation}
- \bar{w}_{1}^{(0)} \ \frac{16 \sqrt{2}}{f} \ \langle \phi \ d v \ \square \, d v \rangle .
\end{equation}

From the extension to the $U(3)$ framework there arise several more terms 
which reduce after applying the above mentioned methods
to the relevant contribution
\begin{equation}
\bar{w}_{2}^{(1)} \,  \frac{16}{f} \ \eta_0\ \langle d v \ \square \, d v \rangle .
\end{equation}

Furthermore there are terms of order $\mathcal{O}(p^6)$ that contain the chiral symmetry 
breaking object $\chi = 2 B \mathcal{M}$ with $\mathcal{M}$ being the quark mass matrix
\begin{equation}
\begin{array}{rccclccl}
\Lagr_{\chi}^{(6)} & = & &
i\,\bar{W}_3 & \epsilon^{\mu \nu \alpha \beta} \langle N \tilde{H}_{\mu \nu} \tilde{H}_{\alpha \beta} \rangle 
& + & i\,\bar{W}_4 & \epsilon^{\mu \nu \alpha \beta} \langle N \rangle
               \langle \tilde{H}_{\mu \nu} \tilde{H}_{\alpha \beta} \rangle \\[1ex]
& & + & \bar{W}_5 & \epsilon^{\mu \nu \alpha \beta} \langle M \tilde{H}_{\mu \nu} \tilde{H}_{\alpha \beta} \rangle 
& + & \bar{W}_6 & \epsilon^{\mu \nu \alpha \beta} \langle M \rangle
               \langle \tilde{H}_{\mu \nu} \tilde{H}_{\alpha \beta} \rangle . 
\end{array}
\end{equation}
The first two terms arise from conventional $SU(3)$ theory whereas the last two are due to
the extension to $U(3)$. In the differential form notation the relevant contributions to the decay 
$P \to \gamma^{(*)} \gamma^{(*)}$ are given by the contact terms
\begin{equation}
\begin{array}{rcclcl}
\Lagr_{\chi, \, ct}^{(6)} & = & & 
\bar{w}_{3}^{(0)} \ \frac{32 \sqrt{2}}{f} \langle \phi \,\chi \, dv \, dv \rangle 
& + & \bar{w}_{4}^{(0)}\ \frac{32 \sqrt{2}}{f} \langle \phi \,\chi \rangle \langle dv \, dv \rangle \\[1ex]
& & + & \bar{w}_{5}^{(1)} \, \frac{32}{f} \ \eta_0 \langle \chi \, dv \, dv \rangle 
& + & \bar{w}_{6}^{(1)} \, \frac{32}{f} \ \eta_0 \langle \chi \rangle \langle dv \, dv \rangle .
\end{array}
\end{equation}
Combining all counterterms of order $\mathcal{O}(p^6)$ we obtain
\begin{equation}
\Lagr_{ct}^{(6)} = 
- \bar{w}_{1}^{(0)} \ \frac{16 \sqrt{2}}{f} \ \langle \phi \ d v \ \square \, d v \rangle
+ \bar{w}_{2}^{(1)} \,  \frac{16}{f} \ \eta_0 \ \langle d v \ \square \, d v \rangle
+ \Lagr_{\chi, \, ct}^{(6)} \ .
\end{equation}


\end{document}